\documentclass[twocolumn,preprintnumbers,superscriptaddress,endnote,nofootinbib,aps,prd,floatfix]{revtex4}

\usepackage[utf8]{inputenc}

\usepackage{footmisc,enumerate}
\usepackage{multirow}
\usepackage{subfigure}
\usepackage{amsmath,amssymb}
\usepackage{graphicx}
\usepackage{placeins}
\usepackage{bbm}
\usepackage{xspace,slashed}
\usepackage{hyperref}
\hypersetup{colorlinks=true, citecolor=blue, urlcolor=blue, linkcolor=blue}
\usepackage[normalem]{ulem}

\usepackage[autostyle]{csquotes}

\renewcommand{\vec}[1]{{\bf{#1}}}

\newcommand{\eps}{\epsilon}

\newcommand{\lam}{\lambda}
\newcommand{\gae}{\gamma_E}

\begin{document}

\newcommand{\og}{\ensuremath{\tilde{O}_g}\xspace}
\newcommand{\ot}{\ensuremath{\tilde{O}_t}\xspace}

\providecommand{\abs}[1]{\lvert#1\rvert}

\newcommand{\Znunujets}{(Z\to{\nu\bar{\nu}})+\text{jets}}
\newcommand{\Welnujets}{(W\to{\ell\nu})+\text{jets}}
\newcommand{\Znunujet}{(Z\to{\nu\bar{\nu}})+\text{jet}}
\newcommand{\Welnujet}{(W\to{\ell\nu})+\text{jet}} 

\title{Power meets Precision to explore 
the Symmetric Higgs Portal}

\begin{abstract}
We perform a comprehensive study of
collider aspects of a Higgs portal scenario that is protected by an unbroken ${\mathbb{Z}}_2$ symmetry.
If the mass of the Higgs portal scalar is larger than half the Higgs mass, this scenario becomes
very difficult to detect. We provide a detailed investigation of the model's parameter space 
based on analyses of the direct collider sensitivity at the LHC as well as 
at future lepton and hadron collider concepts and analyse the importance of these searches for this scenario in the context of expected precision Higgs and electroweak measurements. In particular we also consider the associated electroweak oblique corrections
that we obtain in a first dedicated two-loop calculation for comparisons with the potential of, e.g., GigaZ.
The currently available collider projections corroborate an FCC-hh 100 TeV as a very
sensitive tool to search for such a weakly-coupled Higgs sector extension, driven by small statistical uncertainties over
a large range of energy coverage. Crucially, however, this requires good theoretical control. Alternatively, Higgs signal-strength measurements at an optimal FCC-ee sensitivity level could yield comparable constraints.
\end{abstract}

\author{Christoph Englert} \email{christoph.englert@glasgow.ac.uk}
\affiliation{SUPA, School of Physics \& Astronomy, University of Glasgow, Glasgow G12 8QQ, UK\\[0.1cm]}
\author{Joerg Jaeckel} \email{jjaeckel@thphys.uni-heidelberg.de}
\affiliation{Institut f\"ur Theoretische Physik, Universit\"at Heidelberg,
Philosophenweg 16, 69120 Heidelberg, Germany\\[0.1cm]}
\author{Michael Spannowsky} \email{michael.spannowsky@durham.ac.uk}
\affiliation{Institute for Particle Physics Phenomenology, Durham University, Durham DH1 3LE, UK\\[0.1cm]}
\author{Panagiotis Stylianou}\email{p.stylianou.1@research.gla.ac.uk} 
\affiliation{SUPA, School of Physics \& Astronomy, University of Glasgow, Glasgow G12 8QQ, UK\\[0.1cm]}

\preprint{IPPP/20/4}
\pacs{}

\maketitle

\section{Introduction}
\label{sec:intro}
The lack of evidence for new physics beyond the Standard Model (SM) so far observed at the Large Hadron
Collider (LHC)  combined with the requirement of new interactions
to reconcile shortcomings of the SM has motivated a range of new collider concepts that are currently discussed
in the community. With LHC measurements progressing, active discussions are underway to push the energy frontier
with a new hadron machine. This could reach up to 100 TeV centre-of-mass energy
in the case of a Future Circular Collider (FCC) as discussed in case studies~\cite{Abada:2019zxq,Benedikt:2018csr,Abada:2019lih}. The direct discovery potential of such a machine, given its large energy coverage, is apparent
when compared to collider proposals working at smaller energy such as the Compact Linear Collider (CLIC)
or FCC-ee proposals. However, the latter designs typically offer a much more controlled
environment that can be exploited in finding beyond the SM physics through a systematic
deviation in precision data when compared with the SM-expectation. 
A concept that takes this to the extreme is the so-called GigaZ option~\cite{Erler:2000jg,Erler:2001ja,Baer:2013cma} that aims
to revisit $Z$ boson precision physics to push our understanding beyond the constraints obtained with the Large Electron Positron (LEP).

An interesting scenario in this context is the ${\mathbb{Z}}_{2}$-symmetric Higgs portal \cite{Binoth:1996au,Schabinger:2005ei,Patt:2006fw,Ahlers:2008qc,Batell:2009yf,Englert:2011yb} that is parametrised by
the lagrangian
\begin{multline}
\label{eq:model}
{\mathcal{L}}
={\cal{L}}_{\text{SM}} +\frac{1}{2}(\partial_{\mu} S)^2-\frac{m^{2}_{S}}{2}S^2-\lambda S^2(\Phi^{\dagger}\Phi-v^2/2)\,,
\end{multline}
where $\lambda$ specifies the Higgs portal coupling with the SM Higgs doublet $\Phi$. The latter acquires a vacuum expectation value (vev) $|\left\langle \Phi \right\rangle|=v/\sqrt{2}$
around which we expand as follows,
\begin{equation}
\label{eq:hdoub}
\Phi={1\over \sqrt{2}}\left(\begin{matrix} \sqrt{2} G^+ \\ {v+H+iG^0} \end{matrix}\right)\,,
\end{equation}
with the physical Higgs boson $H$ and the would be Goldstones $G$.

The Higgs portal at and above the electroweak scale presents a particularly interesting and relevant challenge for both high precision and high power approaches. 
For new particle masses $\sim {\rm few}\times 100\,{\rm GeV}$ other ``portals'' to a dark sector, such as the kinetic mixing and the neutrino portal, sensitivity to a level corresponding to a loop-effect, characterised by dimensionless couplings in the range $\sim 10^{-3}-10^{-2}$ is in sight, either with already existing or at least with proposed machines (cf.~\cite{Strategy:2019vxc} for a useful summary). In the case of the Higgs portal we are still far away from this level of sensitivity. 

The case of the symmetric Higgs portal symmetry is particularly interesting as the resulting scalar is stable and therefore a potential dark matter candidate~\cite{Silveira:1985rk,McDonald:2001vt,Davoudiasl:2004be,Barger:2007im,Batell:2011tc,Khoze:2014xha,Feng:2014vea,Athron:2018ipf,Arcadi:2019lka,Hardy:2018bph,Bernal:2018kcw,Alonso-Alvarez:2019pfe}. However, because the new scalar can only be pair produced it also provides for additional challenges.\footnote{A significant ${\mathbb{Z}}_{2}$ breaking can change this situation significantly. We remark, however, that even in this case the current sensitivity is quite limited for scalar masses $m_{S}\gtrsim 100~\text{GeV}$.}  In this work we therefore focus on the case of an unbroken ${\mathbb{Z}}_{2}$ symmetry. 

Searching for a new particle that is weakly coupled, quite heavy and that can only be produced in pairs seems to require both power and precision. To seek the optimal combination we therefore perform a detailed sensitivity study of the scenario of Eq.~\eqref{eq:model} at the aforementioned different collider concepts. In particular we contrast the direct sensitivity that can be expected 
at future lepton and hadron machines with the indirect reach of precise $Z$-pole and electroweak measurements, extending previous work~\cite{Noble:2007kk,Craig:2013xia,Kribs:2017znd,Curtin:2014jma,Craig:2014lda,Ruhdorfer:2019utl,He:2016sqr,Voigt:2017vfz,Englert:2019eyl}.
We demonstrate how the different collider concepts can gain sensitivity to the interactions of Eq.~\eqref{eq:model}.

This work is organised as follows. In Sec.~\ref{sec:pairprod}, we discuss collider processes that show direct sensitivity to this scenario at lepton and hadron colliders and outline selection criteria to isolate the new physics signal from contributing backgrounds. Sec.~\ref{sec:svirt} is dedicated to indirect new physics effects, including a discussion of two-loop oblique corrections in Sec.~\ref{sec:oblique}. Our results are presented and discussed in Sec.~\ref{sec:collider} before we summarise and conclude in Sec.~\ref{sec:conc}.

\section{High-energy and Precision implications}
\label{sec:impli}
For $m_S> m_H/2$ the only known access paths to a scalar coupled via the symmetric Higgs portal are through
either off-shell production of scalar pairs via intermediate Higgs states or footprints of virtual $S$ contributions modifying SM correlations. In this section we will detail both effects.

\medskip

Before setting out on the calculation, let us first define our input parameters.
The vacuum expectation value is related to the electroweak measurements via
\begin{equation}
\label{eq:vevdef}
v = {2m_W s_W \over e},
\end{equation}
where $m_W,s_W,e$ are the $W$ boson mass, the sine of the Weinberg angle, and the QED coupling constant $e=\sqrt{4\pi \alpha}$. The fine structure constant $\alpha$ given by
\begin{equation}
\alpha={\sqrt{2}\over \pi} G_F m_W^2 s_W^2={\sqrt{2}\over \pi} G_F m_W^2 \left(1 - {m_W^2\over m_Z^2}\right)
\end{equation}
with $m_Z$ and $G_F$ denoting the $Z$ boson mass and Fermi constant, respectively.

\subsection{Direct Sensitivity: $S$ Pair Production via Off-Shell Higgs Bosons}
\label{sec:pairprod}
This channel shares phenomenological properties with invisible Higgs decays except for the additional Higgs virtuality 
suppression if $m_S>m_H/2$. As such the rates quickly decrease as a function of $m_S>m_H/2$ (see, e.g., ~\cite{Arcadi:2019lka}). Yet sensitivity is still attainable at large luminosities and higher energies where (improved) signal vs. background suppression is traded off against larger statistics. Here we include weak boson fusion (see also \cite{Eboli:2000ze,Craig:2014lda,Ruhdorfer:2019utl}, Higgs boson-associated gauge boson production~\cite{Godbole:2003it,Davoudiasl:2004aj} as well as mono-jet~\cite{Feng:2005gj,Englert:2011us,Craig:2014lda,Ruhdorfer:2019utl} signatures taking into account the full $m_t$ dependence. Events are generated using the {\sc{FeynRules}}~\cite{Christensen:2008py,Alloul:2013bka}, {\sc{NloCt}}~\cite{Degrande:2014vpa}, {\sc{MadEvent}}~\cite{Alwall:2011uj,deAquino:2011ub,Alwall:2014hca} toolchain. Events showered with {\sc{Pythia8}}~\cite{Sjostrand:2014zea} in the {\sc{HepMC}} format~\cite{Dobbs:2001ck} are passed to {\sc{Rivet}}~\cite{Buckley:2010ar} for analyses. 

Where possible, we compare our results with existing analyses, in particular~Refs.~\cite{Craig:2014lda,Ruhdorfer:2019utl}, and find very good agreement.

Our analysis strategy for hadron colliders follows existing ATLAS and CMS searches, relaxing the missing energy selections in light of the suppressed off-shell signal rate. Search strategies at lepton colliders typically follow similar selections with modifications that we detail below.

\subsubsection*{Hadron Colliders}
\begin{figure}[t]
    \includegraphics[width=8.3cm]{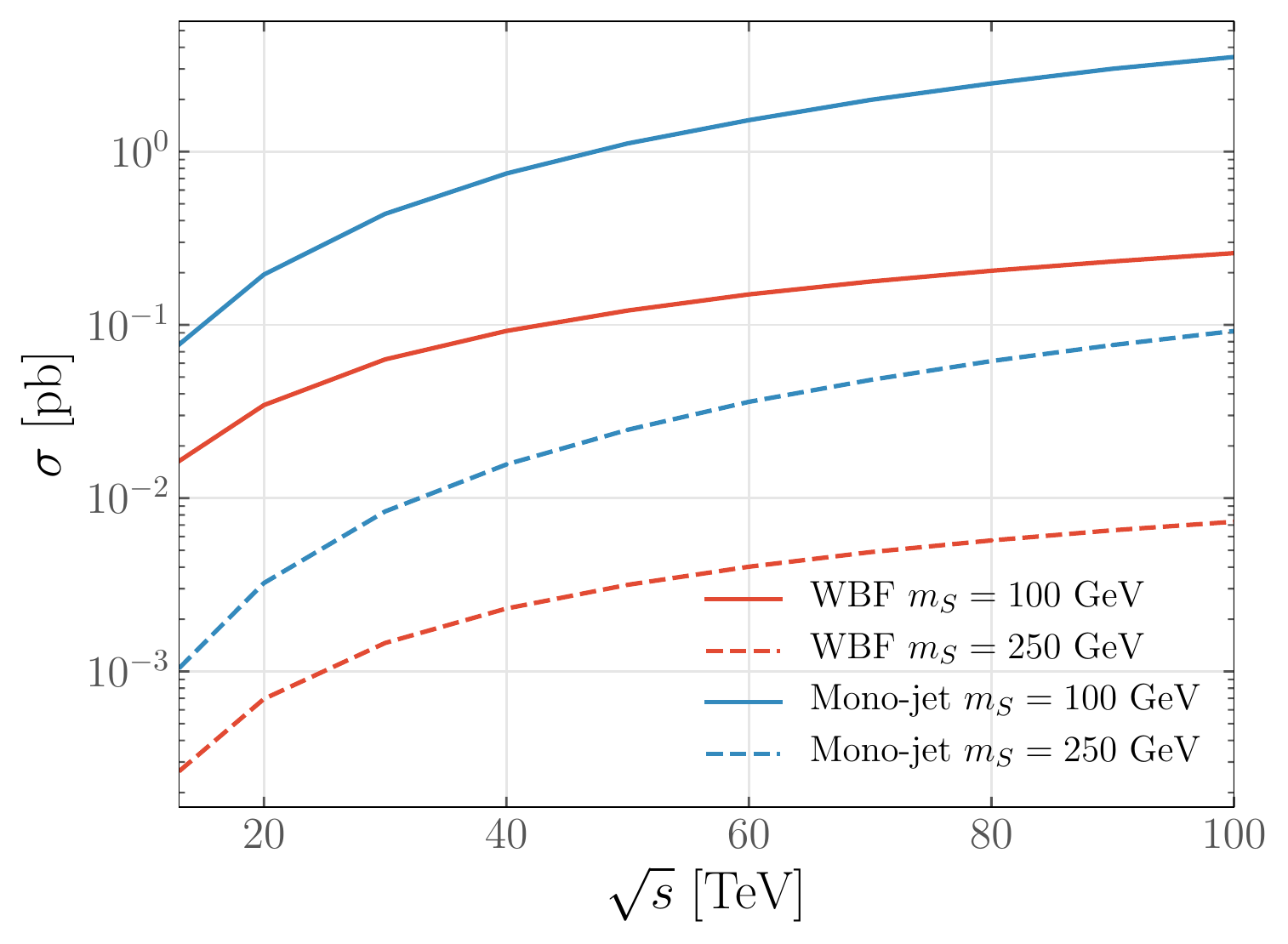}
    \caption{Total cross section values for WBF production of an $S$ pair and two jets through an off-shell Higgs boson at different hadron collider energies (red). We also show the   mono-jet signature (blue), mostly originating from gluon fusion. The associated $(H\to SS)Z$ production is suppressed by 3 orders of magnitude compared to WBF production and therefore not included.} 
    \label{fig:hadronxsec}
\end{figure}

As already mentioned, production of a pair of scalars typically occurs via an off-shell Higgs. Therefore, we consider the channels analogous to those of Higgs production. At hadron colliders we consider three channels involving the pair production of the new scalar: associate production, weak boson fusion and a mono-jet channel resulting mostly from gluon fusion.
The resulting cross sections for $S$ pair production as a function of the centre of mass energy in proton-proton collisions are shown in Fig.~\ref{fig:hadronxsec}. See also \cite{Ruhdorfer:2019utl,Craig:2014lda} for previous analyses of weak boson fusion and mono-jet signatures.

\bigskip

\noindent \underline{{Associate production:}}
We find that $S$ pair production through the associate Higgs production modes is highly suppressed at hadron colliders. For instance, for 100~TeV proton-proton collisions, and using $\lambda=1$ and a relatively light $m_S\simeq 100$~GeV we obtain a signal cross section of ${\cal{O}}(10^{-2})$ fb before any cuts. This is a too small cross section to be phenomenologically relevant in the light of expected backgrounds and uncertainties. It is therefore reasonable to not include associated production $pp \to Z (H\to SS)$ in our comparison.

\bigskip 

\begin{figure}[h]
    \includegraphics[width=8.3cm]{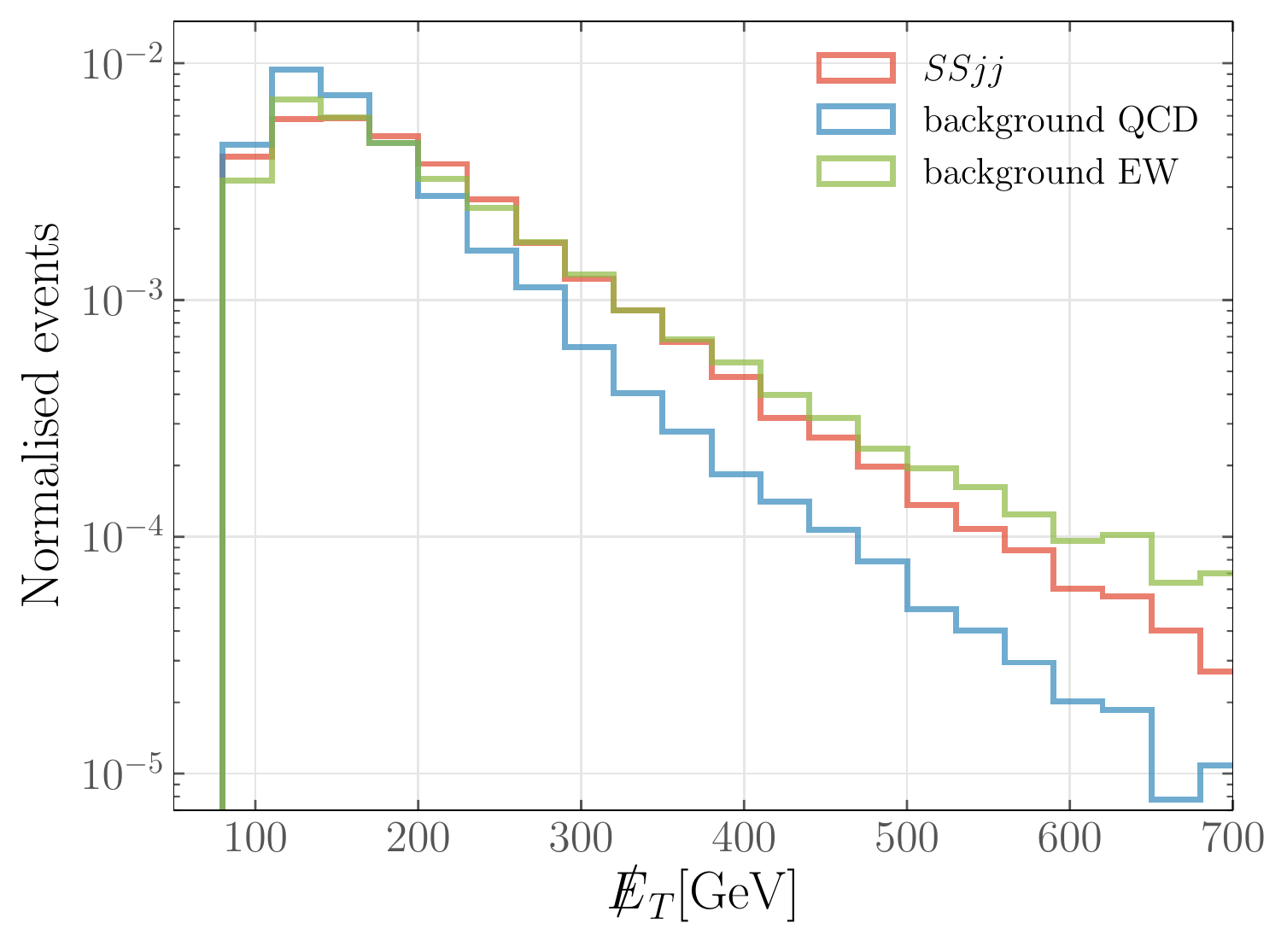}
    \caption{Normalised signal and background distributions of $\slashed{E}_T$ for $m_S = 100$ GeV and $\lambda = 1$ at $100$ TeV FCC. The distributions are obtained with the baseline cuts given in the caption of Tab.~\ref{tab:wbfhadron}. 
    }
    \label{fig:wbfhadronetmiss}
\end{figure}

\begin{table*}[!t]
    \centering
\begin{tabular}{lccccccc}
\hline
    Cuts                  &   $S S j j\;\text{[pb]}\quad$ &   $Z j j\;\text{[pb]}\quad$ &   $W^+ j j\;\text{[pb]}\quad$ &   $W^- j j\;\text{[pb]}\quad$ &   $Z j j\;\text{EW [pb]}\quad$&   $W^+ j j\;\text{EW [pb]}\quad$&   $W^- j j\;\text{EW [pb]}\quad$\\
\hline
 Baseline              &  0.0238  & 10.103   &    6.6287   &   3.0501   &     0.9386 &       0.5897 &       0.3833 \\
 $\Delta \eta > 4.2$   &  0.0217  &  6.6052  &    4.4727  &   1.9775   &     0.8325 &       0.5232 &       0.3384 \\
 $\slashed{E}_T > 200$~GeV &  0.0080 &  1.5842  &    0.7633 &   0.2666  &     0.3952 &       0.1668 &       0.0940  \\
 $m_{jj} > 2300$~GeV       &  0.0041  &  0.3637 &    0.2409 &   0.0637 &     0.2256 &       0.1071 &       0.0594 \\
\hline
\end{tabular}\newline
    \caption{$S$ pair production and background cross sections for WBF at $100$ TeV FCC. The $S$ parameters are set to $m_S = 100$~GeV and $\lambda = 1$, while baseline cuts denote the cuts described in section \ref{sec:pairprod}, but with the relaxed restrictions $\Delta \eta _{jj} > 4.0$, $\slashed{E}_T > 100$~GeV and $m_{jj} > 800$~GeV. QCD corrections will not change these estimates as shown in Ref.~\cite{Mangano:2016jyj}.}  
    \label{tab:wbfhadron}
\end{table*}

\noindent \underline{{Weak boson fusion:}}
Events with $S$ particles generated through weak boson fusion (WBF) are contaminated by $\Znunujets$ and $\Welnujets$ processes, with jets originating from either strong or weak interactions. The WBF signal is characterised by a large pseudorapidity $\eta$ separation of high invariant-mass (back-to-back) tagging jets~\cite{Barger:1991ib,Rainwater:1999sd}. We cluster jets with the anti-kT algorithm \cite{Cacciari:2008gp} with size $0.4$ following Ref.~\cite{Sirunyan:2018owy} and select events with two jets satisfying $p_T(j)\geq 50~\text{GeV}$ in the region of the hadronic calorimeter parametrised by pseudorapidities $|\eta(j)|<4.7$. Enforcing the WBF signal topology, we require a large pseudorapidity separation of the tagging jets $\abs{\Delta\eta({jj})} > 4.0$ at small azimuthal angle $\abs{\Delta\phi_{jj}} < 1.5$ while the jets are required to lie in opposite detector hemispheres $\eta(j_1)\eta(j_2) < 0$. We impose a central jet veto~\cite{Barger:1994zq} to suppress QCD-induced signal and background processes by requiring no jets above $p_T\geq 30$~GeV between the tagging jets. Given these requirements, top pair production as well as QCD multi-jet production do not constitute dominant backgrounds~(see, e.g., \cite{Eboli:2000ze}).

The $\Welnujets$ contamination is further reduced by vetoing events with isolated leptons.\footnote{For the electrons (muons) we define isolation as the sum of $p_T$ of all particle candidates inside a cone of radius $R = \sqrt{(\Delta \eta )^2 + (\Delta \phi )^2} = 0.3 (0.4)$. If the isolation is less than $16~(25) \%$ of the electron (muon) $p_T$, then the lepton is considered isolated \cite{Sirunyan:2018owy}.} To minimise the impact of jet energy scale uncertainties we further require the azimuthal angle difference between the missing transverse momentum vector $\slashed{\vec{p}}_T$ and the transverse momentum of each jet $\vec{p}_T$ must be greater than $0.5$ rad. This criterion is only applied on jets with $p_T > 30~\text{GeV}$ and is considered sufficient to remove multi-jet production. We will refer to the aforementioned cuts as baseline cuts.

With these cuts
the missing energy distribution of signal and background is plotted in Fig.~\ref{fig:wbfhadronetmiss}.
From this we define our search region at sizeable missing energy $\slashed{E}_T =  \abs{\vec{p}_T}> 200$ GeV and require the invariant mass of the leading jets, $m_{jj} > 2.3$ TeV.
For an example with $m_{S}=100~\text{GeV}$ the effects of the cuts on signal and background are given explicitly in Tab.~\ref{tab:wbfhadron}.

\bigskip

\noindent \underline{{Mono-jet production:}}
An $S$ pair can also be produced with an additional jet, through next-to-leading order (NLO) processes that lead to a single Higgs boson recoiling against QCD radiation. 
Selection of events is done by requiring a leading jet of $p_T > 30$ GeV and $\abs{\eta} < 2.4$. Radiation of a sub-leading jet with $p_T$ above $30$ GeV is allowed, as long as the azimuthal separation between the two jets satisfies  $\abs{\Delta \phi(j_1, j_2)} < 2.5$, to suppress the dijet events. Contamination in such events occurs from processes that yield a $\Znunujet$ or $\Welnujet$ final state and is reduced by vetoing any events with isolated electrons or muons. Top and QCD production are subdominant backgrounds~(e.g.~\cite{Aad:2011xw}). Considering the above as the baseline cuts of our analysis, the missing transverse energy is subsequently restricted to $\slashed{E}_T > 150$ GeV and the leading jet transverse momentum to $p_T > 100$ GeV.

\begin{figure}[!t]
    \includegraphics[width=8.3cm]{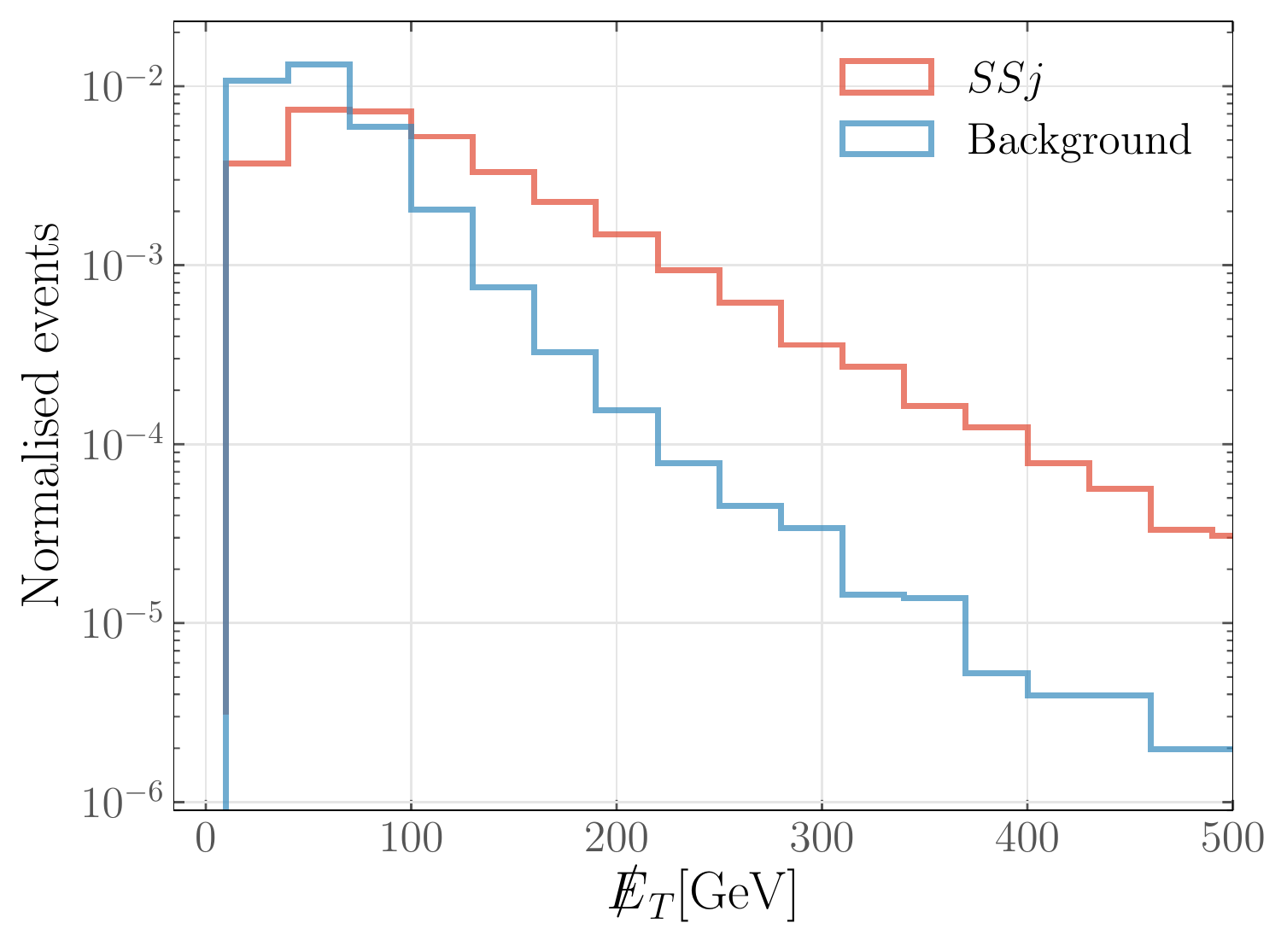}
    \caption{Mono-jet and $S$ pair $\slashed{E}_T$ distribution for $m_S = 100$ GeV and $\lambda = 1$ at $100$ TeV FCC along with combined background using the baseline cuts specified in the caption of Tab.~\ref{tab:monojet}.}
    \label{fig:monojetetmiss}
\end{figure}

Also for this case we compare signal and background as a function of the missing energy in Fig.~\ref{fig:monojetetmiss}. The effects of the cuts are demonstrated for the same example as above in Tab.~\ref{tab:monojet}.\footnote{We also include the subdominant non-gluonic partonic processes not discussed in \cite{Craig:2014lda} and use the transverse mass of the Higgs boson, instead of the partonic center-of-mass energy. This leads to a slight increase in cross section compared to \cite{Craig:2014lda} rendering gluon fusion slightly more sensitive in our comparison. It furthermore highlights the relevance of theoretical uncertainties for all these analyses, an issue that we will not further touch upon in this work.}

\begin{table*}[!t]
\begin{center}
\begin{tabular}{lcccc}
\hline
    Cuts                        &   $S S j\; \text{[pb]}\;\;$  &   $Z j \; \text{[pb]}\;\;$ &   $W^- j\; \text{[pb]}\;\;$&   $W^+ j\;\text{[pb]}$ \\
\hline
 Baseline                    & 0.9322 & 15283   & 17495    & 19799  \\
 $p_T(j_1) > 100$ GeV        & 0.2858 & 820.54  & 553.20   & 670.02 \\
 $\slashed{E}_{T} > 150$ GeV & 0.1810 & 298.28  & 87.381   & 138.12 \\
\hline
\end{tabular}
\caption{ Cross sections for the production of an $S$ pair and a monojet event at $100$ TeV FCC, with $m_S = 100$~GeV and $\lambda = 1$. Background events are also displayed and for the baseline cuts, no $\slashed{E}_T$ restriction is enforced and a relaxed $p_T(j_1) > 30$ GeV is required. $W$ background events are generated with with a minimum lepton cut of $\eta > 2.5$ to enhance statistics. Contamination from $t \bar{t} j$ was significantly smaller than the rest of the background processes and is therefore not included. We take into account
    approximate QCD corrections to the backgrounds via the global $K\simeq 1.6$ factors reported in \cite{Mangano:2016jyj}.} 
    \label{tab:monojet}
    \end{center}
\end{table*}

\subsubsection*{Lepton Colliders}
\begin{figure}[!t]
    \includegraphics[width=8.3cm]{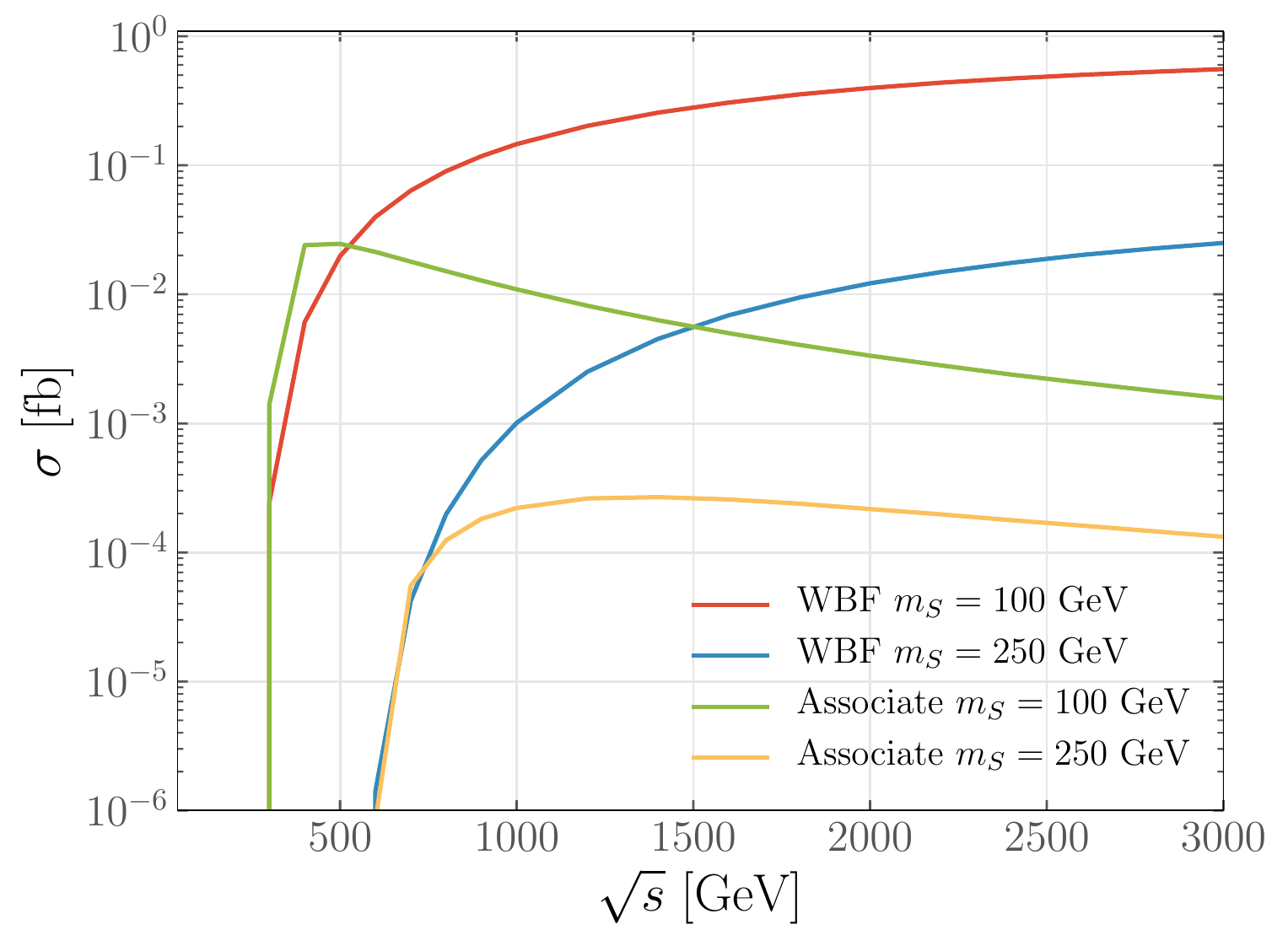}
    \caption{Total cross section values for WBF and associate production of $S$ pair and two leptons at different energies.} 
    \label{fig:leptonxsec}
\end{figure}

In analogy to what we have done for the case of hadron colliders we consider for lepton colliders the two main channels for scalar pair production via an off-shell Higgs: associate production and weak boson fusion (see in particular \cite{Ruhdorfer:2019utl} for a recent analysis). For illustration we show an example of the cross section as a function of the centre of mass energy in Fig.~\ref{fig:leptonxsec}. The events for the cross sections, as well as the rest of the analysis, are generated with the requirements $p_T^\ell > 10$~GeV, $\abs{\eta(\ell)} < 5$ and $\Delta R _{\ell \ell} > 0.4$ applied on light leptons $\ell$.

\bigskip

\noindent \underline{{Associate production:}}
In contrast to hadron colliders, associate Higgs production at lepton colliders is relevant and comparable to the WBF modes. The signal process $e^- e^+ \rightarrow Z (H \rightarrow S S)$, where the on-shell $Z$ boson decays to a lepton pair, is contaminated by $e^- e^+ \rightarrow \ell^- \ell^+ \nu_\ell \bar{\nu}_\ell$, where the neutrinos appear as missing energy. The signal is characterised by a smaller pseudorapidity separation between the lepton pair ($\Delta \eta_{\ell\ell}$) and thus the search region is restricted to $\Delta \eta_{\ell\ell} < 1.3$. Further distinction from the background is achieved with a cut on the missing energy, $\slashed{E}_T > 150$ GeV, and on the missing invariant mass
\begin{equation}
\text{MIM} = \sqrt{\slashed{p}_\mu \slashed{p}^\mu} \geq 200~\text{GeV}\,,
\end{equation}
 where $\slashed{p} = (\sqrt{s}, \vec{0}) - p_{\ell^-} - p_{\ell^+}$.

\begin{figure}[!b]
    \includegraphics[width=8.3cm]{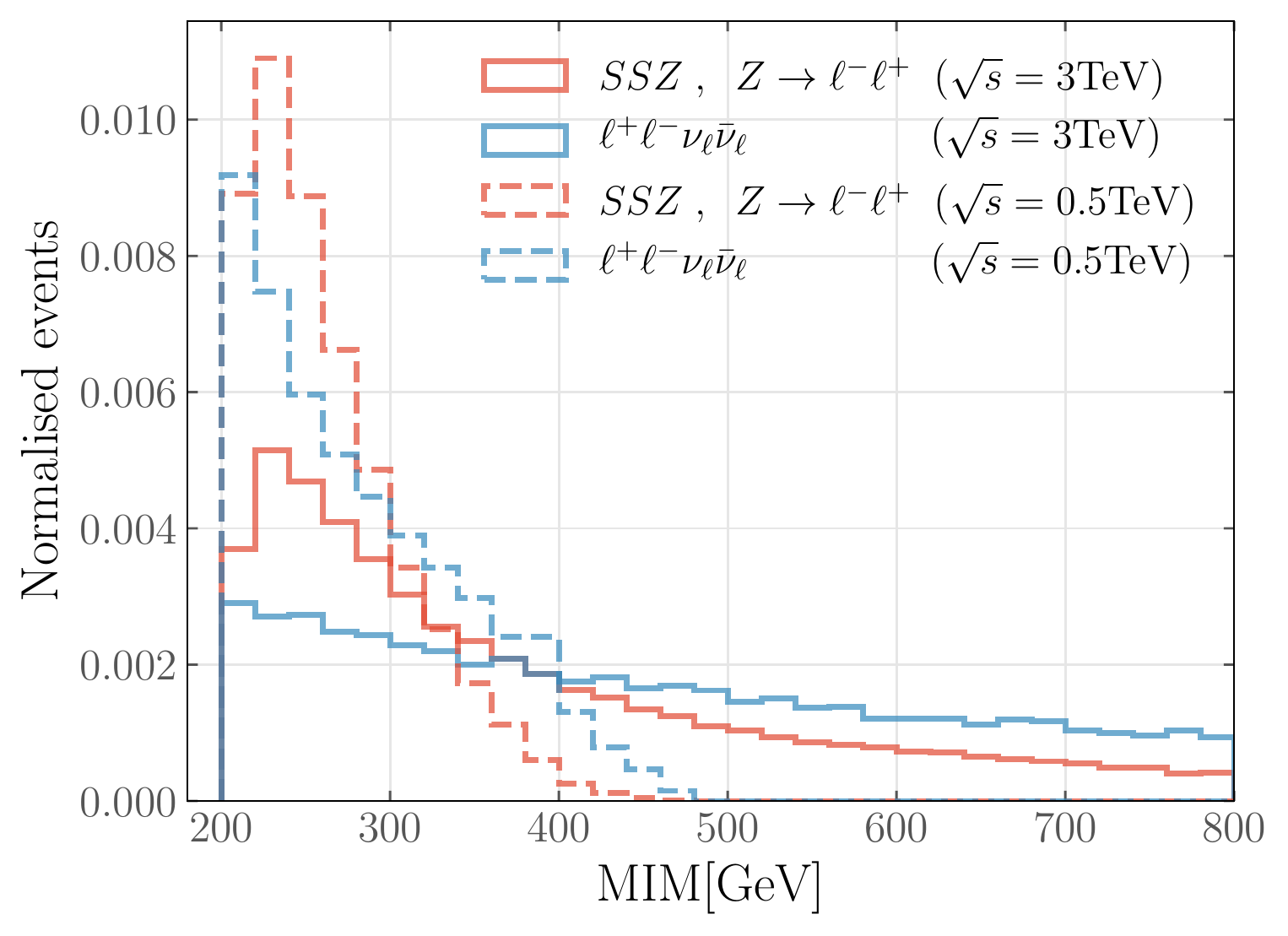}
    \caption{Distributions of MIM for signal and background events with  $m_S = 100$ GeV and $\lambda = 1$ at $500$ GeV ILC, where the associate production will be more relevant, as well as $3$~TeV CLIC.}
    \label{fig:associatemim}
\end{figure}

An example of the MIM distribution of signal and backgrounds is shown in Fig.~\ref{fig:associatemim}. The effects of the cuts are demonstrated in Tab.~\ref{tab:associate}.

\begin{table}[!b]
    \centering
\begin{tabular}{lcc}
\hline
    Cuts                      &   $S S Z , Z \rightarrow \ell^+ \ell^-\; \text{[fb]}\quad$ &   $\ell^+ \ell^- \nu_{\ell} \bar{\nu}_{\ell\;}\text{[fb]}$ \\
\hline
 Generation                   &                         0.0236 &                               669.68   \\
 $\Delta \eta _{\ell\ell} < 1.3$ &                         0.0194 &                               139.64   \\
$\slashed{E}_T > 150$~GeV      &                       0.0113 &                                13.786  \\
	$\text{MIM} > 200$~GeV               &                         0.0113 &                                 2.8209 \\
 $M_{\ell\ell} < 120$~GeV            &                         0.0113 &                                 2.3947  \\
\hline
\end{tabular}\newline
    \caption{ Cross sections for the associate production of an $S$ pair at lepton colliders with $\sqrt{s} = 500$ GeV. Parameters were set to $m_S = 100$~GeV and $\lambda = 1$.} 
    \label{tab:associate}
\end{table}

\bigskip 

\begin{figure}[!t]
    \includegraphics[width=8.3cm]{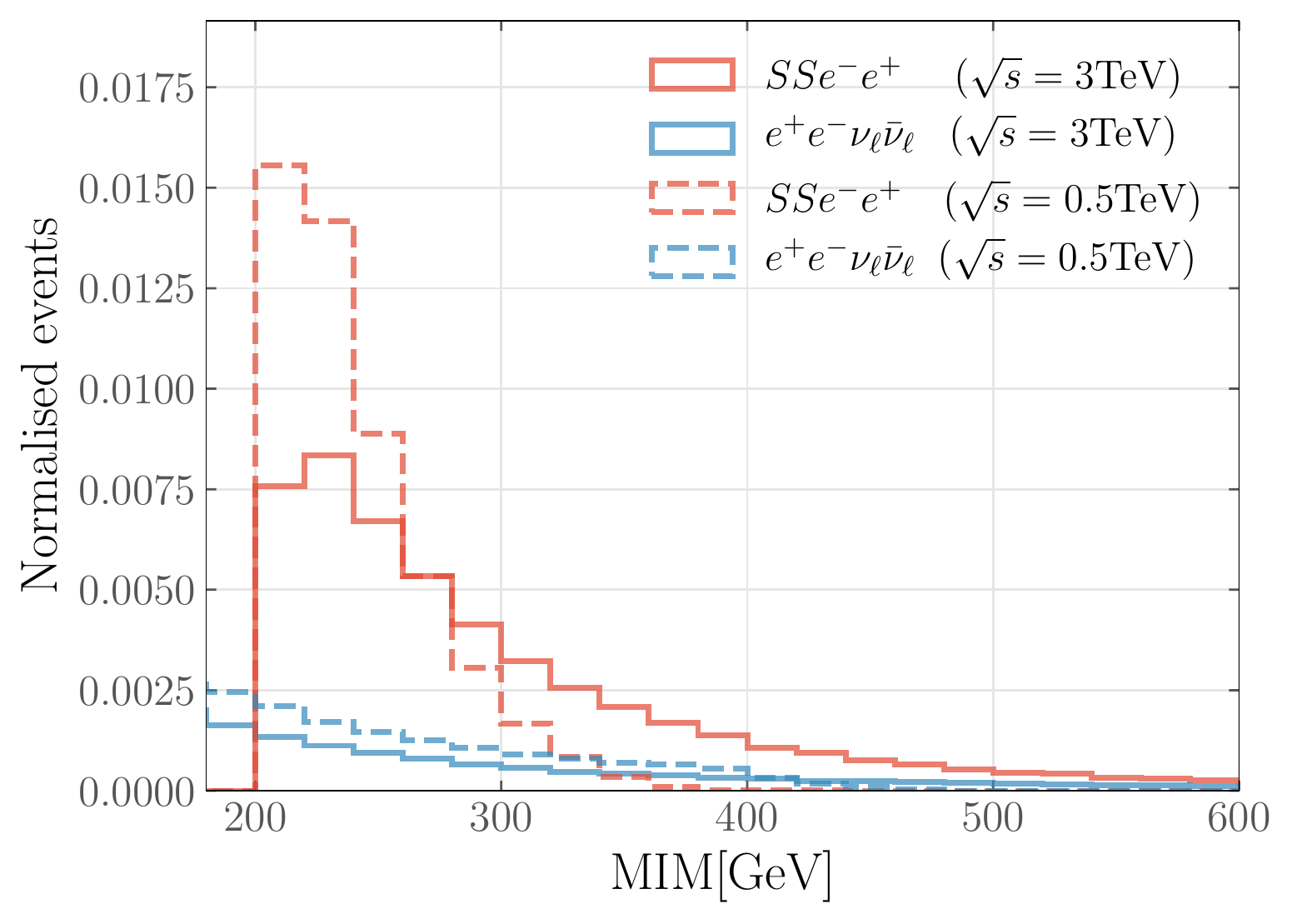}
    \caption{Normalised signal and background distributions of MIM for $m_S = 100$ GeV and $\lambda = 1$ at 500 GeV and 3 TeV lepton colliders produced through WBF. Generation level cuts as in Tab.~\ref{tab:wbflepton} were used.}
    \label{fig:wbfleptonmim}
\end{figure}

\noindent \underline{{Weak boson fusion:}}
WBF remains the dominant process at centre of mass energies larger than $500$ GeV and it is essential to distinguish it from the associate production. This can be achieved with a cut on the invariant mass of the electron-positron pair. For CLIC at $3$ TeV we use $M_{ee} > 2200$~GeV to isolate the signal from contributing backgrounds. The background is further reduced with cuts on the same quantities as in the associate production case. $\Delta \eta_{ee} > 6$ is imposed, since WBF results in leptons with large pseudorapidity separation. The search region is restricted to $\slashed{E}_T > 80$ GeV and MIM $ > 200$ GeV.
For ILC at $500$ GeV the relaxed restrictions $M_{ee} > 120$~GeV and $\Delta \eta_{ee} > 2.0$ were used and the rest of the cuts were kept the same. The former cut also removes any signal event produced via associated modes.
Examples of the MIM distribution and the cutflow are given in Fig.~\ref{fig:wbfleptonmim} and Tab.~\ref{tab:wbflepton}.

\begin{table}[!t]
    \centering
\begin{tabular}{lcc}
\hline
    Cuts                    &   $S S e^- e^+ \;\; \text{[fb]}\quad $&   $e^+ e^- \nu_{\ell} \bar{\nu}_{\ell}\; \text{[fb]}$ \\
\hline
 Generation                &        0.5364 &                               43.86   \\
 $\text{MIM} > 200$~GeV     &        0.5364 &                                9.257 \\
 $\Delta \eta _{ee} > 6$    &        0.4144 &                                1.687 \\
 $\slashed{E}_T > 80$~GeV   &        0.2811 &                                1.446 \\
 $M_{ee} > 2200$~GeV        &        0.2346 &                                0.468 \\
\hline
\end{tabular}   
    \caption{ Cross sections for the $S$ pair and background production for WBF at $3$ TeV CLIC, with $m_S = 100$~GeV and $\lambda = 1$. Cuts are enforced at generation level to improve statistics and include a cut on the sum of neutrino's momenta $\slashed{E}_T^{\nu} > 70$~GeV as well as requiring an invariant electron mass of $M_{ee} > 1500$~GeV. The latter also removes any event arising from associate production.} 
    \label{tab:wbflepton}
\end{table}
\bigskip

Finally, in a WBF topology, where $W$ bosons fuse to produce the Higgs (and neutrinos from the electron and positron), one could use initial state radiation emitted from the colliding electrons or mediating $W$ bosons to trigger the event. In this case, the final state would consist of only a photon and missing energy ($S$ pair and neutrinos) and background contamination would arise from $e^{+} e^{-} \rightarrow \gamma \nu_e \bar{\nu}_e$.  After generating relevant events, we found a significance $N_S / \sqrt{N_B} = 0.0082$, where $N_S$ and $N_B$ are signal and background events respectively. Hence, this is not an avenue to significantly gain sensitivity to the hidden scalars.

\subsection{Indirect Sensitivity: Virtual $S$ imprints}
\label{sec:svirt}
Let us now turn to the indirect measurements, where $S$ is only present in loops (see~\cite{Noble:2007kk,Craig:2013xia,Curtin:2014jma,Craig:2014lda,He:2016sqr,Kribs:2017znd, Voigt:2017vfz,Englert:2019eyl} for previous studies using such observables). Here, we will consider precision observables that are measured at both hadron and lepton colliders. The discussion therefore applies to both types of colliders.

\bigskip

The interactions of Eq.~\eqref{eq:model} will create corrections to the Higgs and Goldstone boson two-point function. The Higgs potential  contained in ${\cal{L}}_{\text{SM}}$ is
\begin{multline}
\label{eq:tadpole}
V (\Phi) = \mu^2 |\Phi|^2 + \lambda_H  |\Phi |^4 \supset v(\mu^2 +v^2\lambda_H)H  = t H\,.
\end{multline}
At leading order this is minimised through conveniently choosing
\begin{equation}
t=v(\mu^2 +v^2\lambda_H)=0\,.
\end{equation}
This choice leads to tadpole diagrams that parametrise the shift of the classical Higgs field value away from the minimum of the Higgs potential as determined by the theory's free parameters 
beyond leading order. In general, Higgs boson tadpoles can be removed from higher order corrections by choosing $t=0$ for bare quantities. This introduces a counterterm $\delta t= - \Gamma^H(p^2=0) $ that corresponds to a renormalisation of the 1-PI Higgs vertex function $\Gamma^H(p^2)$ involving all tadpole diagrams and a correlated Goldstone mass renormalisation (see~\cite{Denner:1991kt,Denner:2019vbn})
\begin{equation}
\delta m_G^2 = - {\delta t \over v} = -{e\over 2 m_W s_W} \delta t 
\end{equation}
The Goldstone renormalisation will be relevant for the discussion of oblique electroweak corrections in Sec.~\ref{sec:oblique}. Note that at one-loop order we can understand $\delta t$ also as
\begin{equation}
\delta v  =  - {\delta t \over m_H^2}
\end{equation}
which shows that working with the ``correct'' vacuum expectation value  in spontaneously broken gauge theories involves tadpole contributions for vertices that result from setting the Higgs to its vev connected by a zero-momentum propagator. As the trilinear Higgs boson interaction vertex follows from the four-point vertex with one leg set to the Higgs' vacuum expectation, the tadpole renormalisation together with the Higgs mass and wavefunction renormalisation constants are also relevant for the corrections to Higgs pair production in Sec.~\ref{sec:hpair}, see~\cite{He:2016sqr,Kribs:2017znd,Voigt:2017vfz,Englert:2019eyl}.

\subsubsection{Higgs coupling modifications}
\label{sec:higgscoup}
Measurements of Higgs boson rates are typically reported using the narrow width approximation owing to the narrowness of the Higgs boson $\Gamma_H/m_H\simeq {\cal{O}}({10^{-5}})$. Signal strengths $\mu$ are then obtained by comparing observations against the SM expectation
\begin{equation}
\label{eq:signal}
\mu = {\sigma(H) \times {\text{BR}} \over [\sigma(H) \times {\text{BR}}]_{\text{SM}}}
\end{equation}
where $\sigma(H)$, ${\text{BR}}$ represent particular Higgs boson production and decay branching modes.
For the model given in Eq.~\eqref{eq:model} when $m_S> m_H/2$ no non-SM Higgs decay channel are present. In this case, all modifications away from the SM will be due to virtual $S$ effects (see Ref.~\cite{Craig:2013xia,Craig:2014lda,Englert:2019eyl} for earlier analyses).

The Higgs wave function and mass squared renormalisation constants in the on-shell scheme are given by
\begin{equation}
\delta Z_H= -{\lam^2 \over 8\pi^2 }{ 2 m_W s_W \over e} \hbox{Re} {\partial B_0(q^2, m_S^2, m_S^2) \over \partial q^2}\bigg| _{q^2=m_H^2}\,,
\end{equation}
and
\begin{multline}
\delta m_H^2={\lam^2 \over 8\pi^2} {4 m_W^2 s_W^2 \over e^2} \hbox{Re} B_0(m_H^2,m_S^2,m_S^2) 
\\+{\lam\over 16\pi^2} \hbox{Re} A_0(m_S^2) 
\end{multline}
with Passarino-Veltman~\cite{Passarino:1978jh} functions $A_0,B_0$ which are given in $D$-dimensional regularisation in e.g.~Ref.~\cite{Denner:1991kt} (see also \cite{Scharf:1993ds,Martin:2005qm}). The
$D\to 4$ divergent pieces of the $B_0$ are momentum-independent. This renders $\delta Z_H$ finite for the scenario in this paper and at the given order in perturbation theory. Any single Higgs production process or partial decay width $\Gamma_i$ will then obtain an $S$-correction
\begin{equation}
\label{eq:singmod}
{\sigma(H) \over [\sigma(H)]_{\text{SM} }} =  {\Gamma_i \over  [\Gamma_i]_{\text{SM} }} = 1 + \delta Z_H
\end{equation}
which leads to\footnote{The Higgs wave function renormalisation can be understood as effective operator $\sim (\partial_\mu |\Phi|^2)^2$ which leads to identical conclusions.} (see also \cite{Englert:2013tya,Craig:2013xia,Craig:2014una})
\begin{equation}
\label{eq:strength}
\mu = {\sigma(H) \times {\text{BR}} \over [\sigma(H) \times {\text{BR}}]_{\text{SM}}} = 1 + \delta Z_H\,.
\end{equation}
Constraints on the Higgs signal strength \cite{Englert:2017aqb} can therefore be treated analogously to Higgs portal models with a dark vacuum expectation value leading to Higgs coupling modifications proportional to a characteristic Higgs mixing angle, which can be identified with $\delta Z_H$.

Note that given that the Higgs coupling modifications are uniform, all relevant information in the comparison against the SM is contained in the total cross section and, consequently, in the signal strength constraint. 

\subsubsection{Off-Shell Higgs boson probes}
\label{sec:offshell}
A channel that received considerable interest recently in the context of Higgs coupling studies at hadron colliders is the so-called off-shell measurement of $p(g)p(g)\to H \to ZZ \to 4~\text{leptons}$. Due to unitarity cancellations in the absorptive parts of the amplitude linked to $t\bar t\to ZZ$ scattering, the Higgs contributions are non-decoupling for energies above the Higgs resonance~\cite{Kauer:2012hd}. Correlating Higgs off-shell with on-shell $H\to ZZ$ measurements (Eq.~\eqref{eq:signal}) can then be interpreted as an indirect measurement of the Higgs width \cite{Caola:2013yja,Englert:2014aca} under assumptions of how these different kinematic regions are connected~\cite{Englert:2014ffa}. 

In the scenario of Eq.~\eqref{eq:model} at ${\cal{O}}(\lam^2)$, the $gg\to ZZ$ continuum is unchanged while the Higgs contributions receive corrections from the scalar $S$. The modification of the $s$-channel Higgs exchange amplitude ${\cal{M}}$ is given by 
\begin{multline}
\label{eq:hprobmod}
{{\cal{M}}\over {\cal{M}}_{\text{SM}}} - 1 =-{\lam^2 m_W^2 s_W^2 \over 8 \pi^3 \alpha (s-m_H^2) }  \\ \times \left( B_0(s,m_S^2,m_S^2) -\hbox{Re}B_0(m_H^2,m_S^2,m_S^2)  \right)\,.
\end{multline}
Note that the right hand side vanishes when we take the limit $s\to m_H^2$ as expected from the cancellation of vertex and propagator renormalisations when we do not include the finite lifetime of the Higgs boson with an ad-hoc Breit-Wigner distribution. 
Including the modification of the total Higgs decay width according to Eq.~\eqref{eq:singmod} results again in Eq.~\eqref{eq:strength} upon expansion. 

\begin{figure}[!t]
    \includegraphics[width=8.3cm]{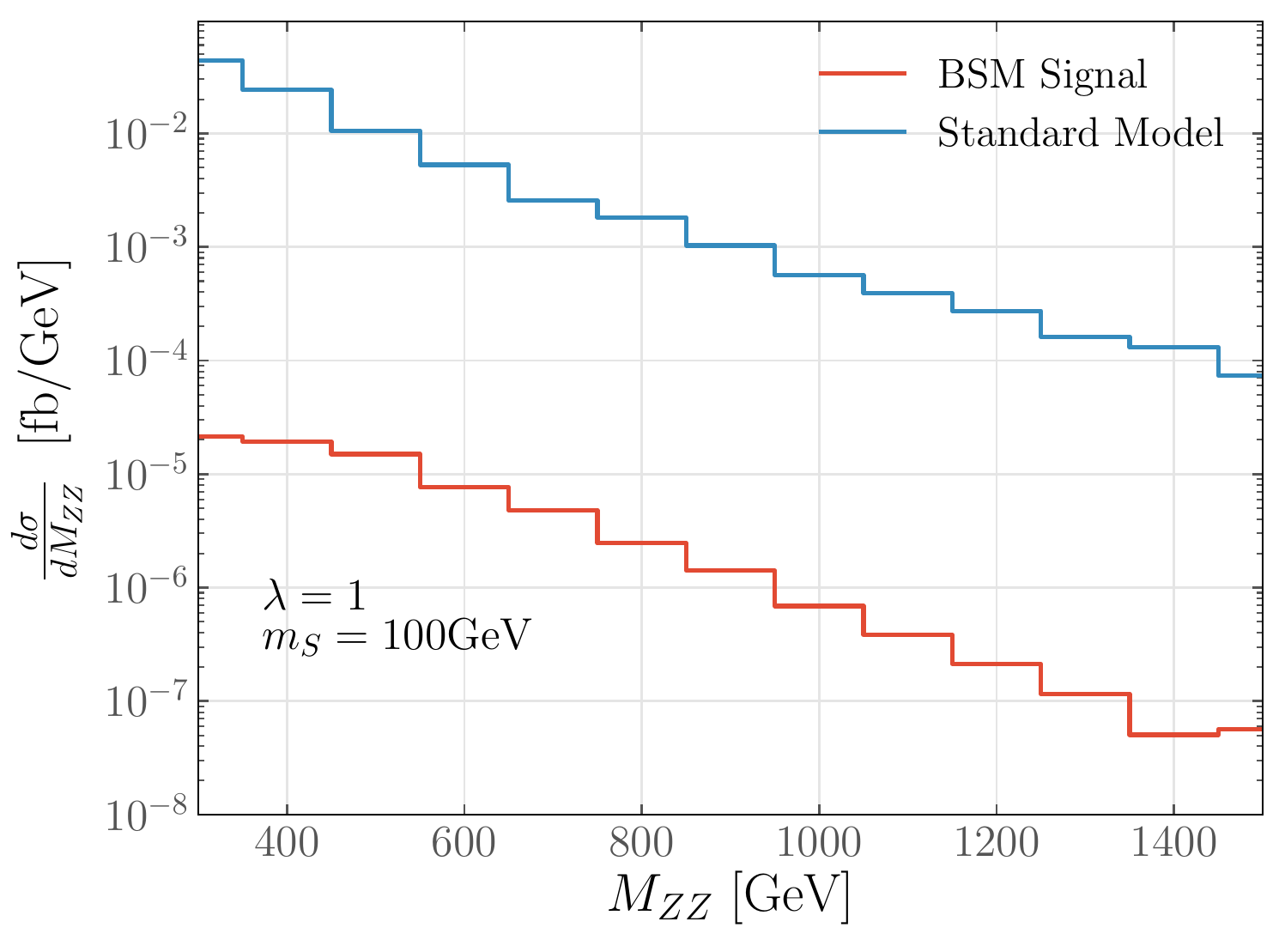}
    \caption{Differential cross sections for $g g \rightarrow Z Z $ at $100$~TeV FCC, indicating that corrections originating from $S$ to the Higgs contribution of $p p \rightarrow 4 \ell$ are negligible. } 
    \label{fig:gg_ZZ}
\end{figure}

This channel only shows limited sensitivity as can be seen from Fig.~\ref{fig:gg_ZZ}. As can be expected from the discussion of Ref.~\cite{Englert:2019zmt}, the corrections of Eq.~\eqref{eq:hprobmod} are small even before interfering with the SM $gg\to ZZ$ continuum amplitude. Even for extrapolations to 30/ab at a 100 TeV FCC-hh that are typically discussed as design targets for such a machine~\cite{Abada:2019zxq,Benedikt:2018csr,Abada:2019lih,Abada:2019ono}, we do not obtain constraints in this channel that are robust in the sense of perturbative unitarity (see below).

\subsubsection{Higgs Pair Production}
\label{sec:hpair}
Virtual $S$-loops also modify Higgs pair production~\cite{Noble:2007kk,Curtin:2014jma,He:2016sqr,Voigt:2017vfz,Englert:2019eyl}.
As the trilinear Higgs boson interaction vertex follows from the four-point vertex with one leg set to the Higgs' vacuum expectation value, the 3-point Higgs function is still a function of the tadpole renormalisation constant $\delta t$ even when we remove tadpoles throughout the calculation by choosing a tadpole renormalisation
\begin{equation}
\label{eq:tren}
\delta t= - {\lam\over 8 \pi^2} {2 m_W s_W\over e}  \hbox{Re}  A_0(m_S^2)\,.
\end{equation}
The amplitude for the relevant $HH$ production (i.e. 
weak boson fusion $e^+e^-\to HH \nu_e\bar{\nu}_e$ at high-energy lepton colliders and $gg\to HH$ at hadron colliders)
is then obtained from expanding the transition probability 
\begin{equation}
|{\cal{M}}|^2 = |{\cal{M}}_{\text{SM}}|^2 + 2\text{Re}\left( {\cal{M}}_{\text{SM}} {\cal{M}}^\ast_{{\lambda}}\right)\,,
\end{equation} 
where SM/$\lambda$ refer to the leading order and next-to-leading order contributions $\sim \lambda$, respectively. We
will consider the next-to-leading correction in the following, see~\cite{Englert:2019eyl}.

\subsubsection{Oblique Corrections}
\label{sec:oblique}

\begin{figure*}[!t]
\parbox{11.5cm}{\includegraphics[width=11.5cm]{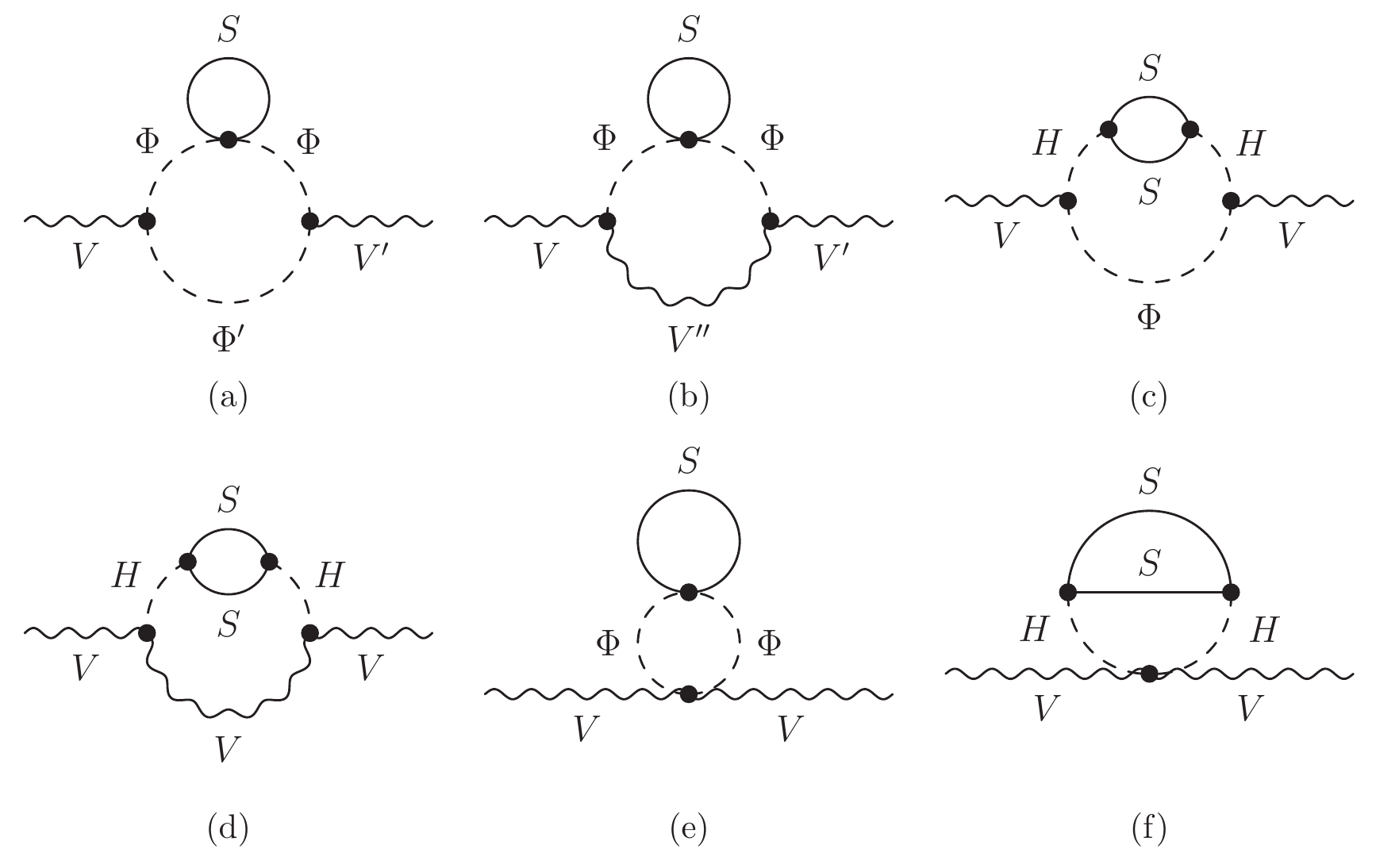}}\hfill
\parbox{5.8cm}{\vspace{3cm}\caption{\label{fig:feynpols} Representative two-loop Feynman diagram topologies of the electroweak boson polarisation functions for boson $V$ that give rise
to the electroweak oblique corrections $S,T,U \sim \lam,\lam^2$. $\Phi,\Phi'$ denote all possible Higgs and Goldstone boson insertions. $V,V',V''=W,Z,A$ label all allowed SM vector boson insertions.}}
\end{figure*}
%
\begin{figure*}[!t]
\parbox{5.8cm}{\vspace{3cm}\caption{\label{fig:feynpolsct} Representative two-loop Feynman diagram counter term topologies of the electroweak boson polarisation functions similar to Fig.~\ref{fig:feynpols}. The first diagram represents two-loop renormalisation constants that are not obtained from one-loop inserted one-loop renormalisation constants. Note that $\Phi\Phi' V''$ vertex counterterms are suppressed.}}
\hfill
\parbox{11.5cm}{\includegraphics[width=11.5cm]{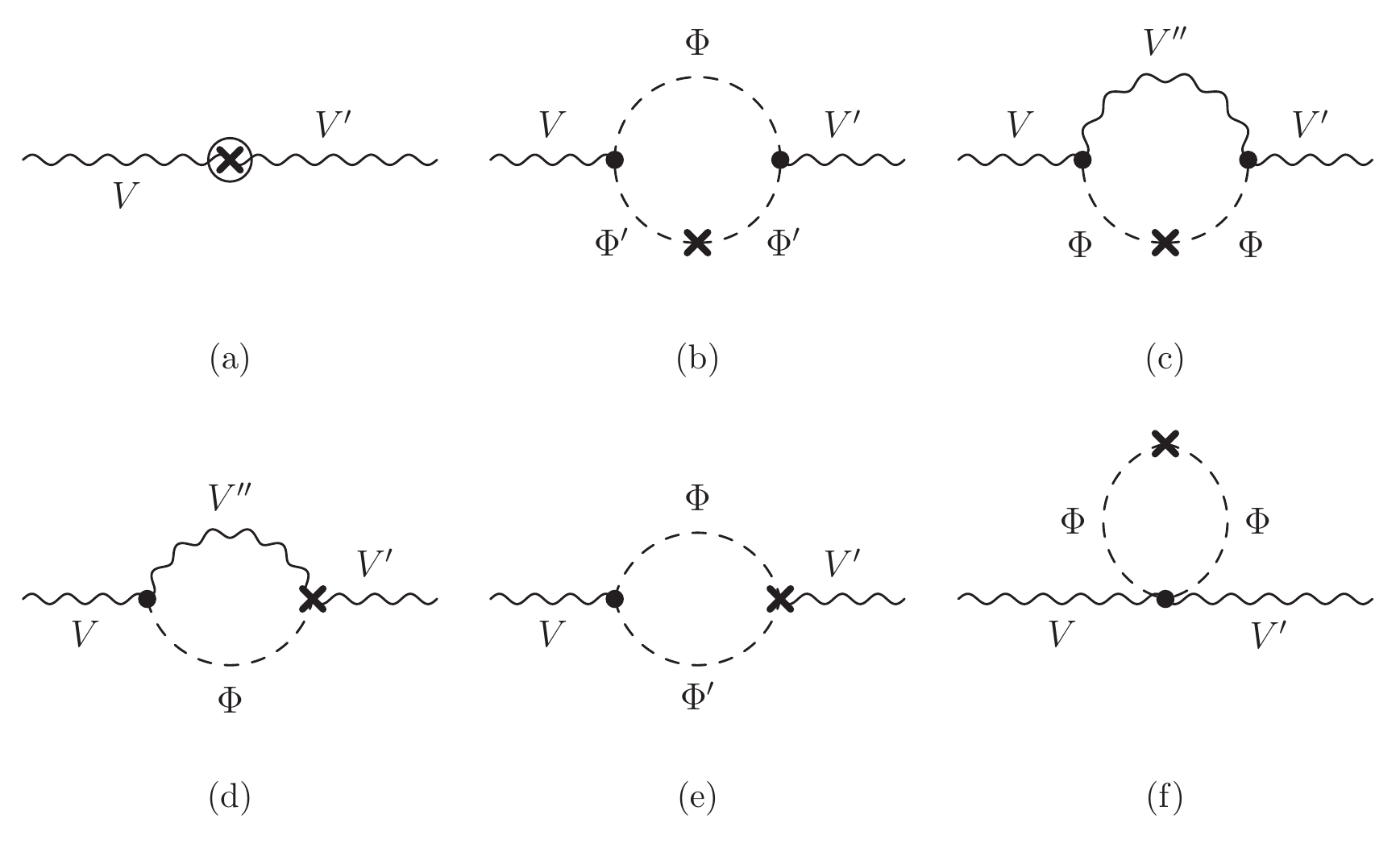}}
\end{figure*}

\newcommand{\zzf}{\Pi_{ZZ}(m_Z^2)}
\newcommand{\zzfn}{\Pi_{ZZ}(0)}
\newcommand{\azf}{\Pi_{AZ}(m_Z^2)}
\newcommand{\azfn}{\Pi_{AZ}(0)}
\newcommand{\aaf}{\Pi_{AA}(m_Z^2)}
\newcommand{\aafn}{\Pi_{AA}(0)}
\newcommand{\wwf}{\Pi_{WW}(m_W^2)}
\newcommand{\wwfn}{\Pi_{WW}(0)}
\newcommand{\alfa}{\alpha}

The Peskin-Takeuchi $S,T,U$ parameters~\cite{Peskin:1990zt} follow from an investigation of polarisation functions
$\Pi_{VV'}$ 
\begin{multline}
\parbox{3.5cm}{\vspace{0.15cm}\includegraphics[width=3.5cm]{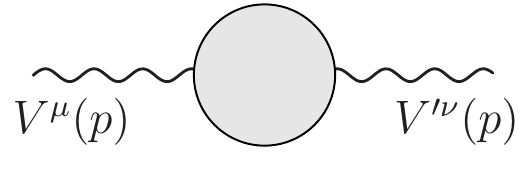}} \sim
\Pi_{VV'}^{\mu\nu}(p^2) \\= 
(p^2-m_V^2) \delta_{VV'} +
 \Pi_{VV'} (p^2) \left( g^{\mu\nu} - {p^{\mu}p^{\nu} \over p^2} \right)\\ +  B_{VV'}(p^2) {p^{\mu}p^{\nu} \over p^2}\,,
\end{multline}
and their transverse parts in particular. The so-called oblique corrections
are then given by (see also~\cite{Golden:1990ig,Holdom:1990tc,Altarelli:1990zd,Grinstein:1991cd,Peskin:1991sw,Altarelli:1991fk,Burgess:1993mg,Barbieri:2004qk})
\begin{widetext}
  \begin{equation}
    \label{eq:stu}
    \begin{split}
      S&= {4 s_W^2c_W^2 \over \alfa} \left(\frac{\zzf - \zzfn}{m_Z^2}
        - {c_W^2-s_W^2\over c_Ws_W}{\azf-\azfn\over m_Z^2} -
        {\aaf\over m_Z^2}\right) \,,\\
      T&= {1\over \alfa}\left( {\wwfn\over m_W^2} - {\zzfn\over m_Z^2}
        - {2 s_W\over c_W} {\azfn\over m_Z^2}\right)\,, \\
      U&= {4 s_W^2\over \alfa} \left( {\wwf-\wwfn\over m_W^2} - c_W^2
        {\zzf - \zzfn\over m_Z^2 } \right.
      -2 s_W c_W
        {\azf-\azfn\over m_Z^2}- \left. s_W^2{\aaf \over m_Z^2}  \right) \,,
    \end{split}
  \end{equation}
\end{widetext}
where $c_W,s_W$ are the cosine and sine of the Weinberg angle, respectively. $S,T,U$ parametrise the leading modifications of gauge boson interactions due to presence of new physics affecting their propagation, i.e. they capture correlated modifications away from the SM expectation of electroweak four-fermion scattering processes.
As the new scalar only couples to the Higgs boson and is protected by the unbroken ${\mathbb{Z}}_{2}$-symmetry, contributions to $S,T,U$ do only arise at two-loop order.
The relevant diagrams and counterterms are given in Fig.~\ref{fig:feynpols} and \ref{fig:feynpolsct}, respectively.

In the definition of Eq.~\eqref{eq:stu} we have already exploited the Ward identity $\aafn=0$ which means that we will work with on-shell renormalised quantities in the following. For instance, for our scalar $S$ insertions we obtain before renormalisation in $D$-dimensional regularisation and using Feynman gauge, Fig.~\ref{fig:feynpols} (a),(b),(e),
\begin{equation}
\Pi^0_{AA} (0) = -{\alpha (D-4) (D-2) \over 256 \pi^3 m_W^2} \lam A_0(m_S^2) A_0(m_W^2)
\end{equation}
where $A_0$ is the standard function one-loop function (expanding $D=4-\eps< 4$)
\begin{multline}
A_0(x) = x\bigg[ 
{2\over \eps} - \gamma_E  - \log {x\over 4\pi \mu^2}  + 
   1 + \\ {\eps \over 4} \big( (-\gae  - \log{ x\over \mu^2 } + 1)^2 + 1 + {\pi^2\over 6}\bigg)\bigg]\,.
\end{multline}
This yields
\begin{multline}
\Pi^0_{AA} (0) = {\alpha \lam m_S^2 \over 32\pi^3 }  \bigg(
{1\over \eps} -  \gae + \log \left({m_S m_W \over 4\pi\mu^2} \right) - {1\over 2} \bigg) \\ + {\cal{O}}(\eps)\,.
\end{multline}
This cancels identically against the renormalised Goldstone contribution
\begin{equation}
\delta \Pi_{AA} (0) = - { \alpha (D-4) (D-2) \over 32 \pi^2 m_W^2} {e \,\delta t \over m_W s_W} A_0(m_W^2)
\end{equation}
with the one-loop tadpole renormalisation $\delta t$ given in Eq.~\eqref{eq:tren}.

\begin{figure}[!t]
{\includegraphics[width=8.3cm]{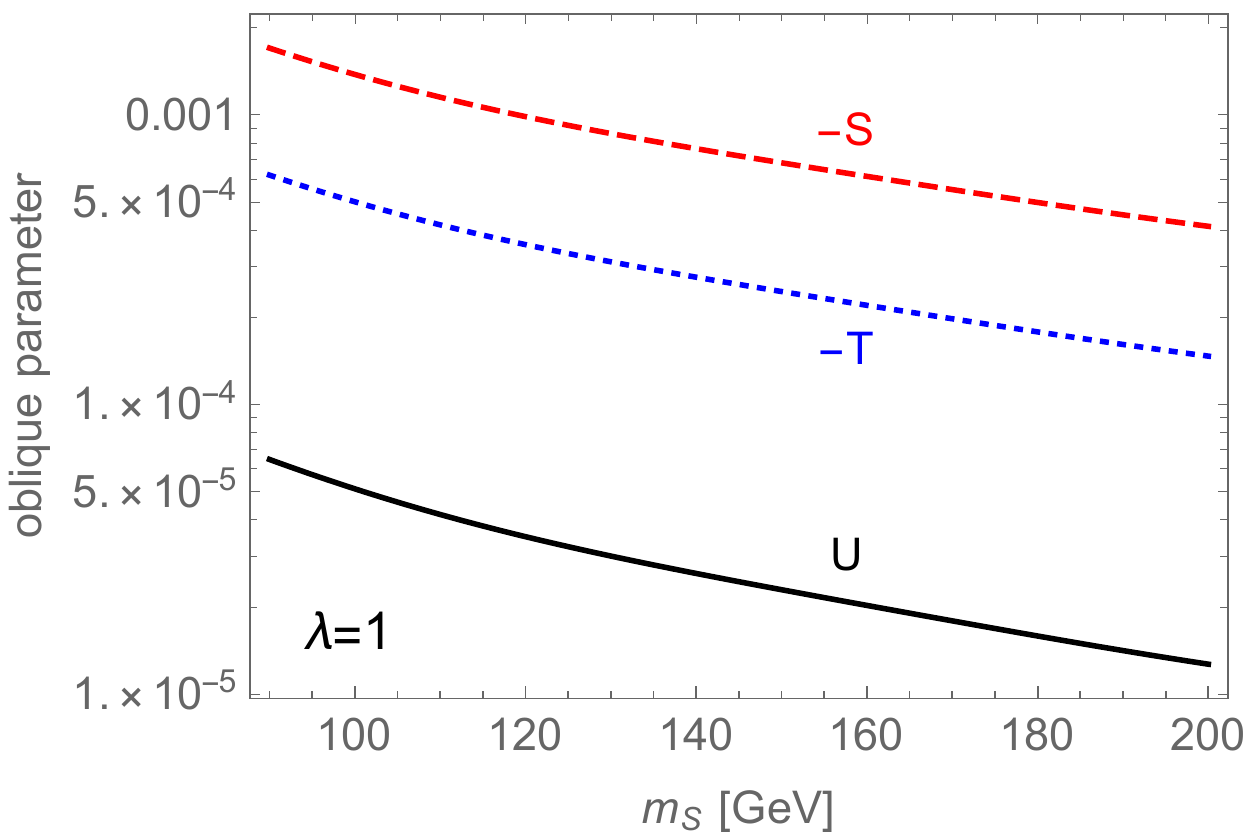}}
\caption{\label{fig:obliqueres} Oblique parameter values for different scalar masses and $\lambda=1$.}
\end{figure}

We compute the oblique corrections using a combination of
{\sc{FeynCalc}}~\cite{Mertig:1990an,Shtabovenko:2016sxi},
{\sc{FeynArts}}~\cite{Hahn:2000kx},
{\sc{LoopTools}}~\cite{Hahn:1998yk,Hahn:2000jm},
{\sc{Tarcer}}~\cite{Mertig:1998vk} and we perform
analytical checks to ensure UV finiteness. Full numerical results
are then obtained by employing {\sc{Tsil}}~\cite{Martin:2005qm}, which is based on Ref.~\cite{Tarasov:1997kx} (see also \cite{Weiglein:1993hd}). 
In Fig.~\ref{fig:obliqueres} the results are shown as a function of the scalar mass.

For the oblique parameters, we note that the $U$ parameter is suppressed by an order of magnitude compared to $S,T$. This can be seen in Fig.~\ref{fig:obliqueres}. This is consistent with the fact that $U$ is not sourced by dimension 6 effective operators. We therefore employ the $U=0$ projections of Ref.~\cite{Baak:2014ora} for the {\sc{GFitter}} LHC 300/fb and ILC/GigaZ options. 

\begin{figure*}[!t]
\subfigure[]{\includegraphics[width=8.3cm]{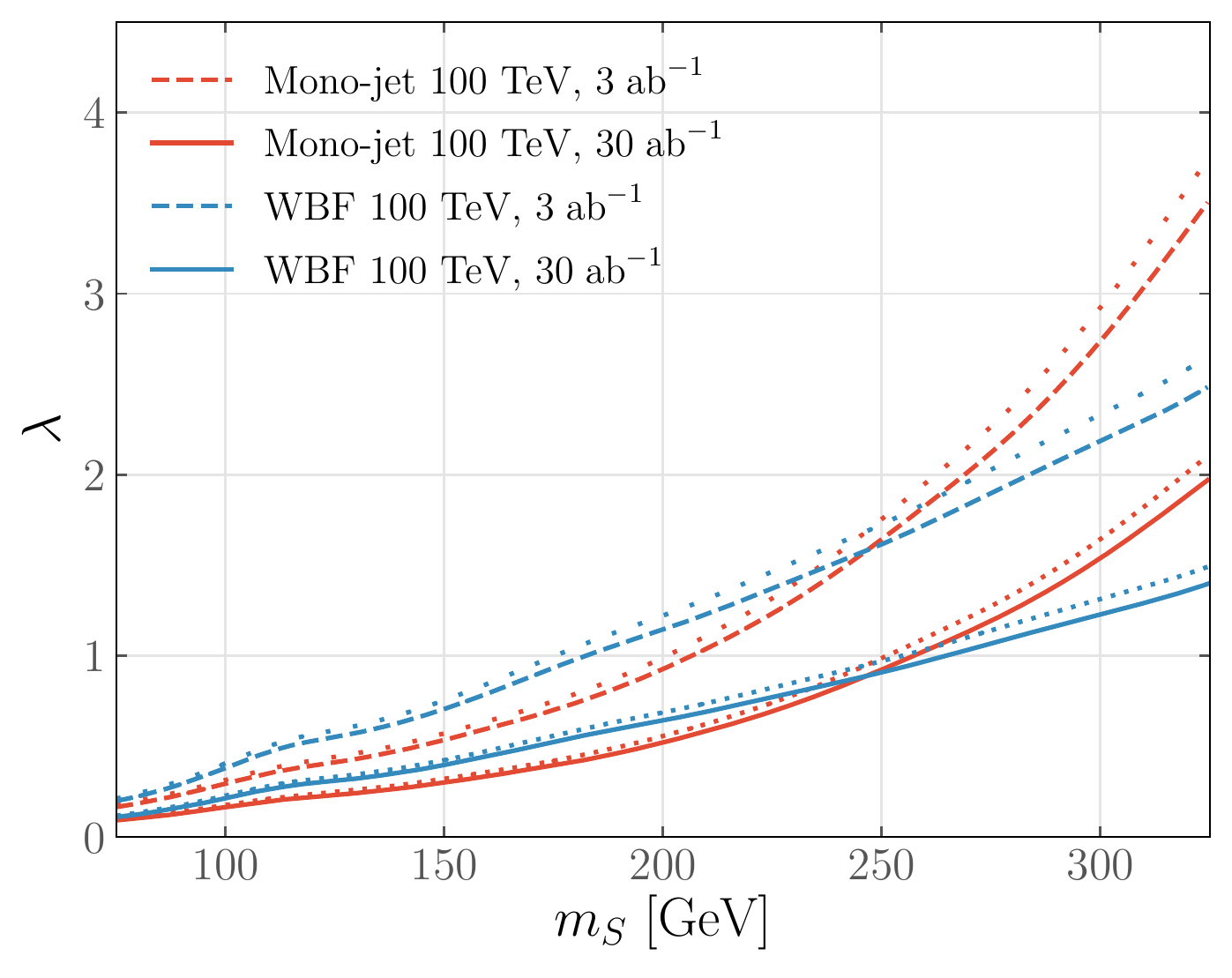}}
\hfill
\subfigure[]{\includegraphics[width=8.3cm]{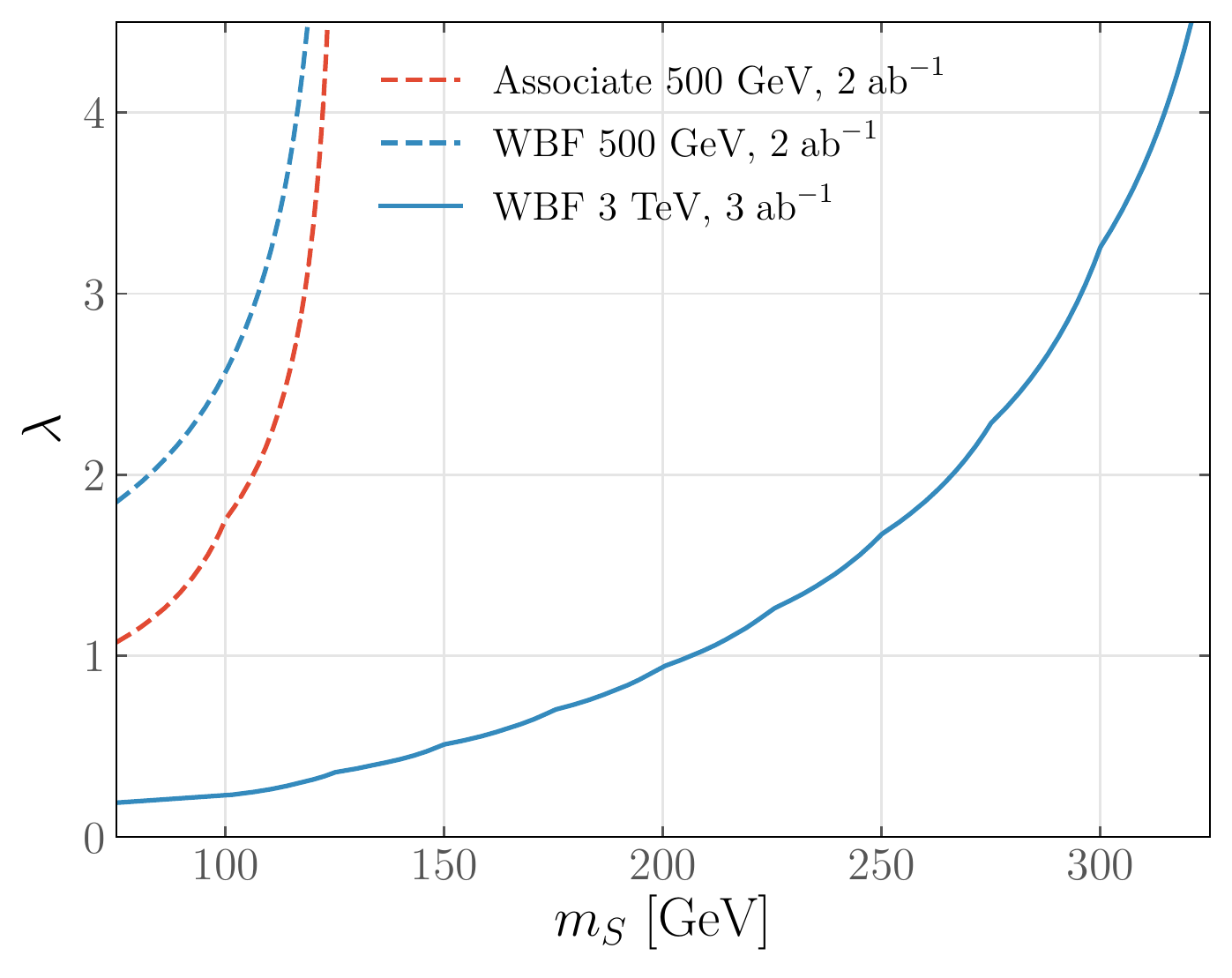}}
    \caption{\label{fig:direct_exclusion} (a) 68\% exclusion regions for FCC at $3\;\text{ab}^{-1}$ and $30\;\text{ab}^{-1}$ integrated luminosities, as identified with $N_S / \sqrt{N_S+N_B} = 1$. Dotted and loosely dotted lines correspond to a $30\%$ increment of the background for the $3\;\text{ab}^{-1}$ and $30\;\text{ab}^{-1}$ cases respectively. (b) 68\% exclusion regions for ILC at $2\;\text{ab}^{-1}$ luminosity and 3 TeV CLIC at $3\;\text{ab}^{-1}$. 
    }
\end{figure*}

\section{Power meets Precision: Expected Collider Sensitivity}
\label{sec:collider}
Before we turn to the discussion of the expected sensitivity to the parameters $\lambda, m_S$ it is instructive
to consider the perturbative unitarity constraints on $\lambda$. Forward $SH\to SH$ scattering in the high energy limit $s=(p_S+p_H)^2\gg m_H,m_S$ and the perturbative constraint on the
zeroth partial wave (see e.g.~\cite{Dawson:2018dcd})
\begin{equation} 
\text{Re}\,a_0 (SH\to SH) \leq{1\over 2}
\end{equation} 
yields straightforwardly $\lambda \lesssim 4\pi$. We find that this limit is quickly
approached at around $\sqrt{s}\simeq 3.5~\text{TeV}$ for the mass range $m_Z< m_S\lesssim 300~\text{GeV}$ that we consider in this work. It is worthwhile to note that this perturbativity constraint is weaker than the electroweak stability bounds, see \cite{Curtin:2014jma}, which limit $|\lambda | \lesssim 1$.

\medskip

The direct cross section measurements are summarised in Fig.~\ref{fig:direct_exclusion}. After all cuts have been applied, the 68\% exclusion regions are obtained by requiring $N_S / \sqrt{N_S+N_B} = 1$, where $N_S$ and $N_B$ are signal and background events, respectively. The analysis was employed for $100$~TeV FCC for integrated luminosities of $3~\text{ab}^{-1}$ and $30~\text{ab}^{-1}$, with WBF being more sensitive at masses larger than $m_S = 250$ GeV compared to mono-jet production. Exclusion regions are studied for $500$ GeV ILC at $2~\text{ab}^{-1}$, which will be reached at later stages of the experiment according to Ref.~\cite{Aihara:2019gcq} and associate production is the most sensitive process in this case. In contrast, WBF is the only significant process at the higher energies of $3$ TeV at $3~\text{ab}^{-1}$ that can be reached by CLIC. 

\medskip
We now turn to the expected precision of Higgs signal strength measurements at different colliders.

Projections of single Higgs measurements in the context of singlet extensions have been provided in, e.g., Ref.~\cite{Klute:2013cx}. We find that the projected global constraints on universal Higgs mixing for 240 GeV lepton colliders are in good agreement with constraints that we obtain from a projection of $e^+e^-\to HZ$ alone.
Based on this we focus on this single measurement to constrain universal Higgs coupling modifications for all types of colliders. 

\begin{figure}[!t]
\subfigure[]{\includegraphics[width=8.3cm]{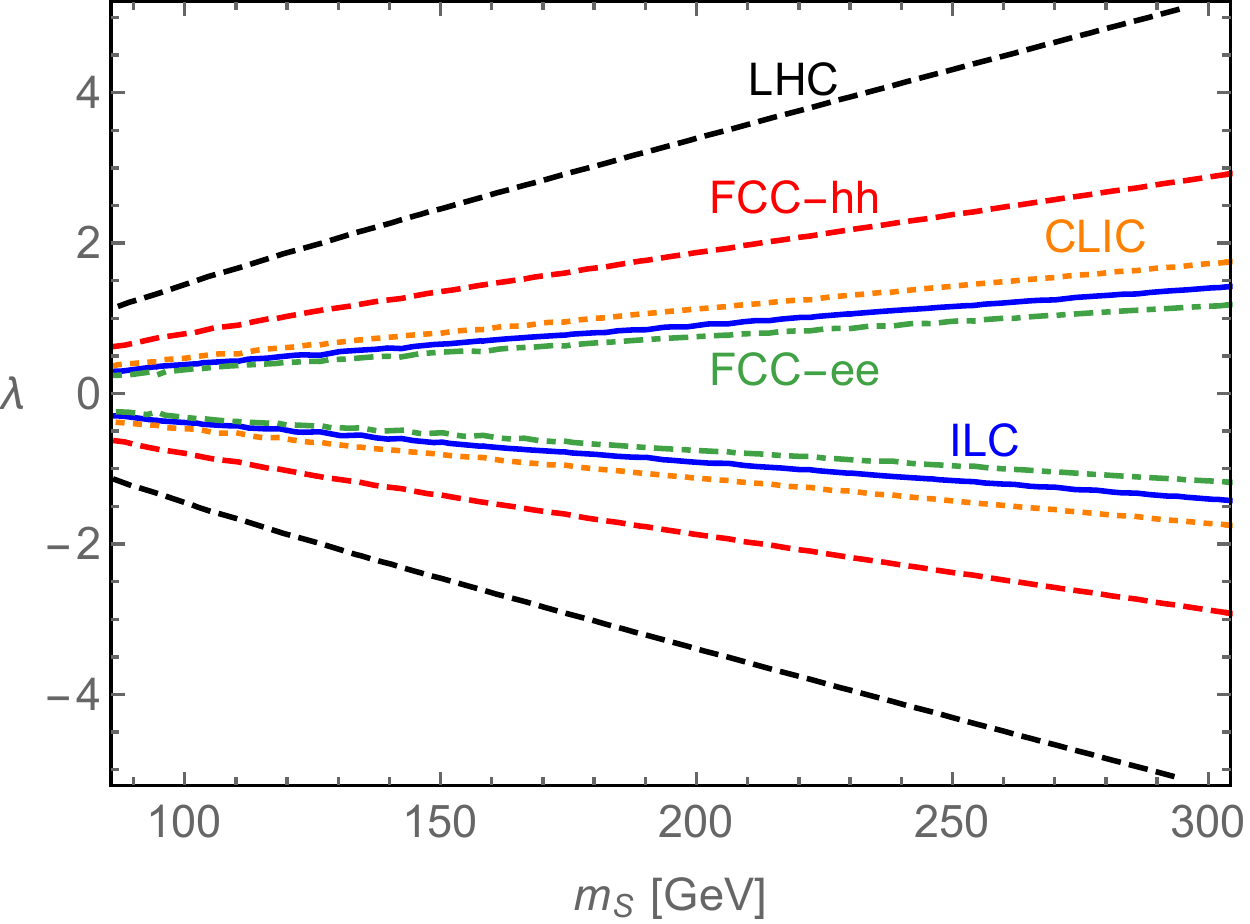}}
\caption{\label{fig:higgssignal} Higgs signal strength measurements at different colliders interpreted in the singlet off-shell model. The limits are based on the uncertainties stated in Eqs.~(\ref{fig:single_higgs})-(\ref{fig:single_higgs_fcchh}).
}
\end{figure}

The fit of Ref.~\cite{Klute:2013cx} gives
\begin{subequations}
\begin{align}
\label{fig:single_higgs}
\text{LHC}: &\quad \mu= [0.96,1.03]
\end{align}
for LHC projections. A dedicated recent analysis of Higgs coupling measurements at lepton colliders~\cite{deBlas:2019rxi} finds
fractional signal strength uncertainty of
\begin{align}
\label{fig:single_higgs_lc}
\text{ILC-250}: &\quad {\delta \mu \over \mu}= 0.29\%\,, \\
\label{fig:single_higgs_lc2}
\text{CLIC-380}: & \quad {\delta \mu \over \mu}=0.44\%\,, \\
\label{fig:single_higgs_lc3}
\text{FCC-ee(240)}: & \quad {\delta \mu \over \mu}=0.2\%\,.
\end{align}
At a future FCC-hh option we can expect \cite{Benedikt:2018csr}
\begin{align}
\label{fig:single_higgs_fcchh}
\text{FCC-hh}: &\quad {\delta \mu \over \mu}= 1.22 - 1.88\%\,,
\end{align}
\end{subequations}
depending on the Higgs decay channel.
The results for the different experiments are
shown in Fig.~\ref{fig:higgssignal}.

\medskip

\begin{figure}[!t]
\includegraphics[width=8.3cm]{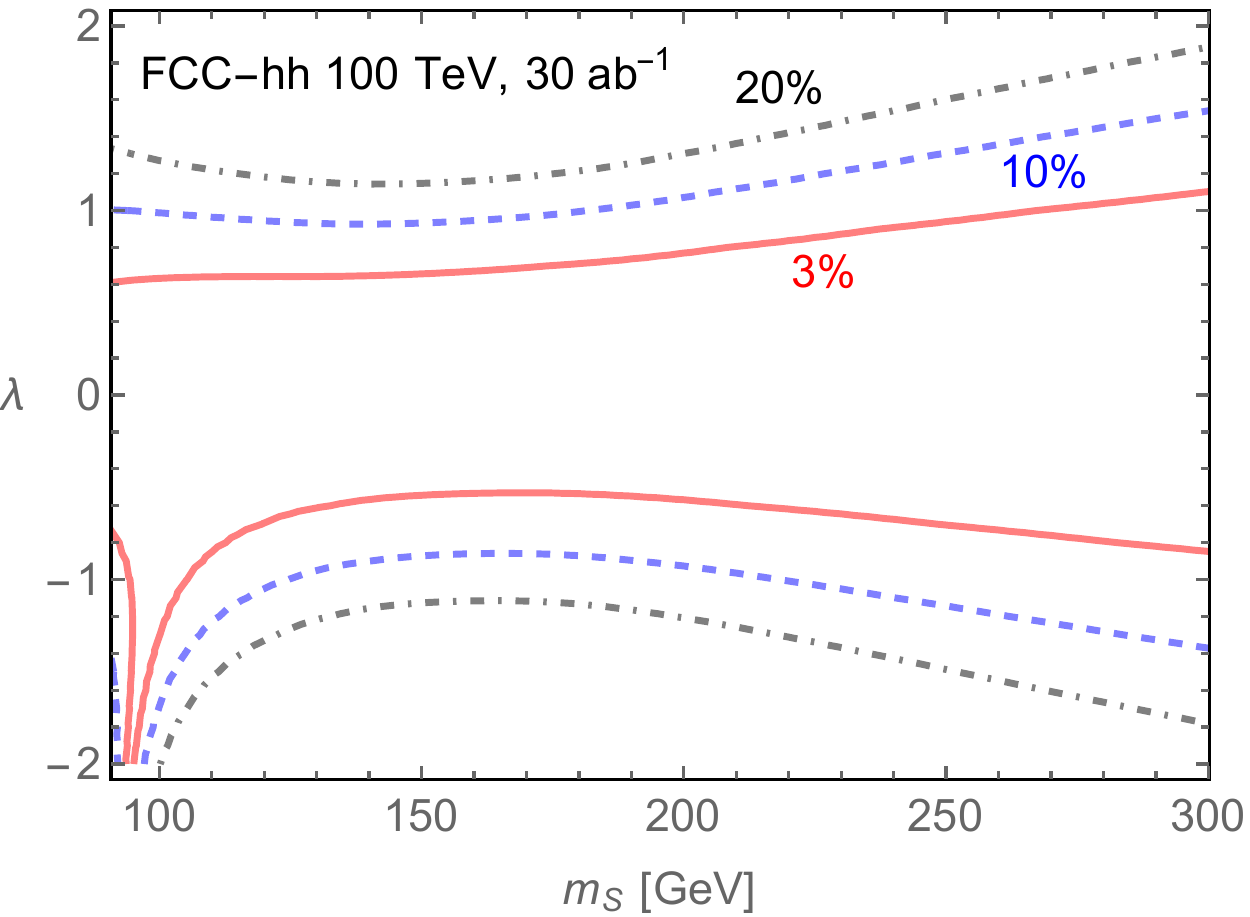}
\caption{\label{fig:hh} Sensitivity of di-Higgs production at the FCC-hh (30/ab) to the symmetric off-shell Higgs portal based on an inclusive $\kappa_\lambda$ measurement. The different lines refer to the expected accuracy of $\kappa_\lambda$, where 3\% is the 68\% confidence level reported in Ref.~\cite{Contino:2016spe}. The di-Higgs results for CLIC are qualitatively identical and given in Fig.~\ref{figlcbc}.}
\end{figure}

\begin{figure}[!t]
\includegraphics[width=8.3cm]{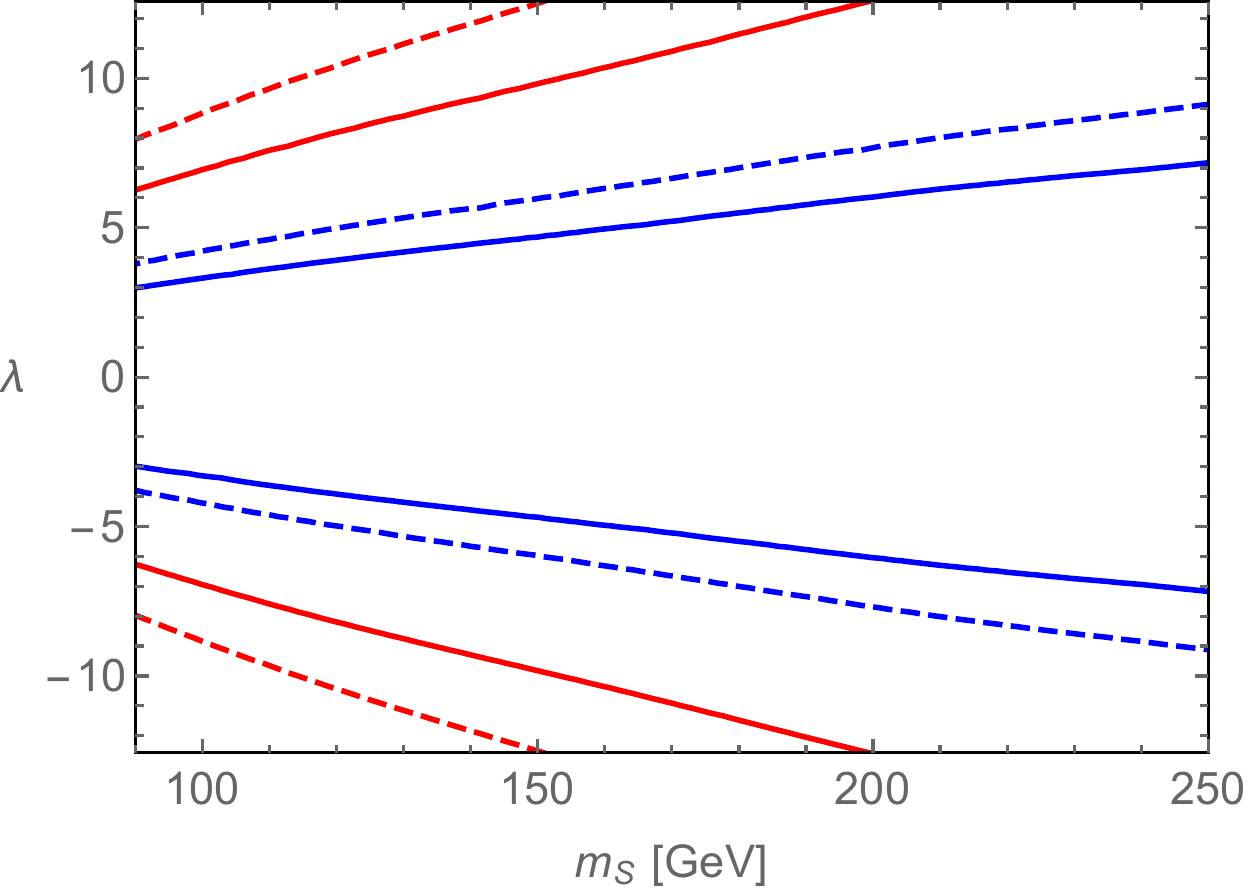}
\caption{\label{fig:stu} (a) LHC 300/fb (red) and ILC/GigaZ (blue) \cite{Baer:2013cma} $S,T$ constraints at 
68\% (solid) and 95\% confidence level (dashed) as provided by {\sc{GFitter}}~\cite{Baak:2014ora} for $U=0$ when considered in the parameter space of the portal model at two-loop level.}
\end{figure}

For measurements of the Higgs self-coupling, a recent CMS projection gives 68\% and 95\% confidence
level projections
$\kappa^{\text{68\%}}_{\lambda_{\text{SM}}}={\lambda^{\text{68\%}}_{\text{SM}} / \lambda_{\text{SM}}  }=  [0.35,1.9]$,
and $
\kappa^{\text{95\%}}_{\lambda_{\text{SM}}}={\lambda^{\text{95\%}}_{\text{SM}} / \lambda_{\text{SM}} }= [- 0.18, 3.6]$ (see Ref.~~\cite{CMS:2018ccd}), where $\lambda_{\text{SM}}$ is the SM Higgs self-coupling. Note that these limits are not much further than a factor of order two away from the perturbative limits of forward $HH$ scattering. While these constraints are perturbatively meaningful they do not suggest large sensitivity to weakly coupled, non-resonant Higgs sector extensions. This is owed to the fact of a relatively small inclusive di-Higgs cross section of about 32~fb at the LHC~\cite{Dawson:1998py,Frederix:2014hta,deFlorian:2015moa,deFlorian:2016uhr,Borowka:2016ehy,Borowka:2016ypz,Heinrich:2017kxx,Grazzini:2018bsd,deFlorian:2018tah,Baglio:2018lrj}. Enhancing sensitivity to Higgs pair production is a key motivation for pushing the energy frontier beyond the LHC. 

Turning to a future hadron collider with 100 TeV centre-of-mass energy, the measurement of the Higgs self-coupling is expected to reach up to $3-6\%$ precision with regard to the machine's and the estimated detector capability~\cite{Contino:2016spe, Banerjee:2018yxy, Banerjee:2019jys}. Following~\cite{Englert:2019eyl} and interpreting inclusive Higgs self-coupling measurements (i.e. the associated cross section difference) of $\kappa_\lambda$ along the lines of $S$-induced corrections, we obtain the results of Fig.~\ref{fig:hh}. We note that 3\% sensitivity is below the currently understood theoretical limitations of $\sim 10\%$~\cite{DiMicco:2019ngk}, which will saturate the uncertainty of the self-coupling measurement extraction unless theoretical improvements become available. For comparison we therefore include re-interpretations of $\delta \kappa_{\lambda_\text{{SM}}}/\kappa_{\lambda_\text{{SM}}}=10\%,20\%$ in Fig.~\ref{fig:hh}. The behaviour of self-coupling measurements in the WBF channel at lepton colliders is qualitatively identical and we will discuss them in the next section.

\medskip

Let us now turn to the oblique parameters. For $S$ and $T$ the correlation matrices, central values and uncertainties for the LHC and GigaZ are given by~\cite{Baak:2014ora}
\begin{align}
\label{fig:corrma}
\text{LHC}: &~\rho= \left( \begin{matrix} 1 & 0.96 \\ 0.96 & 1 \end{matrix} \right) &  (\Delta S,\Delta T) = (0.086, 0.064)\,, \\
\text{GigaZ}: &~\rho= \left( \begin{matrix} 1 & 0.91 \\ 0.91 & 1 \end{matrix} \right) & (\Delta S,\Delta T) = (0.018, 0.023)\,.
\end{align}
From these we can obtain a $\chi^2$ through the inverse error-multiplied correlation matrix, which is translated to our Higgs portal parameters in Fig.~\ref{fig:stu}. As can be seen, the constraints that can be expected at the LHC in the near future are not competitive with the indirect constraints from on-shell Higgs measurements. GigaZ improves this dramatically, however, the sensitivity is still too low for the two-loop contributions to compete with Higgs measurements at Higgs factories such as ILC, CLIC and FCC-ee. An improvement in the electroweak measurement by $\sim 30$ would be necessary to become competitive. While this suggests that electroweak precision measurements are unlikely to play a fundamental role in constraining the parameter space of the outlined off-shell Higgs portal, the fact that the required improvement is much smaller than the naive loop factor suppression $\sim 16\pi^2$ highlights the generic relevance of electroweak precision constraints for general future BSM investigations.

\begin{figure*}[!t]
\subfigure[\label{fig:fccbc}]{\includegraphics[width=8.3cm]{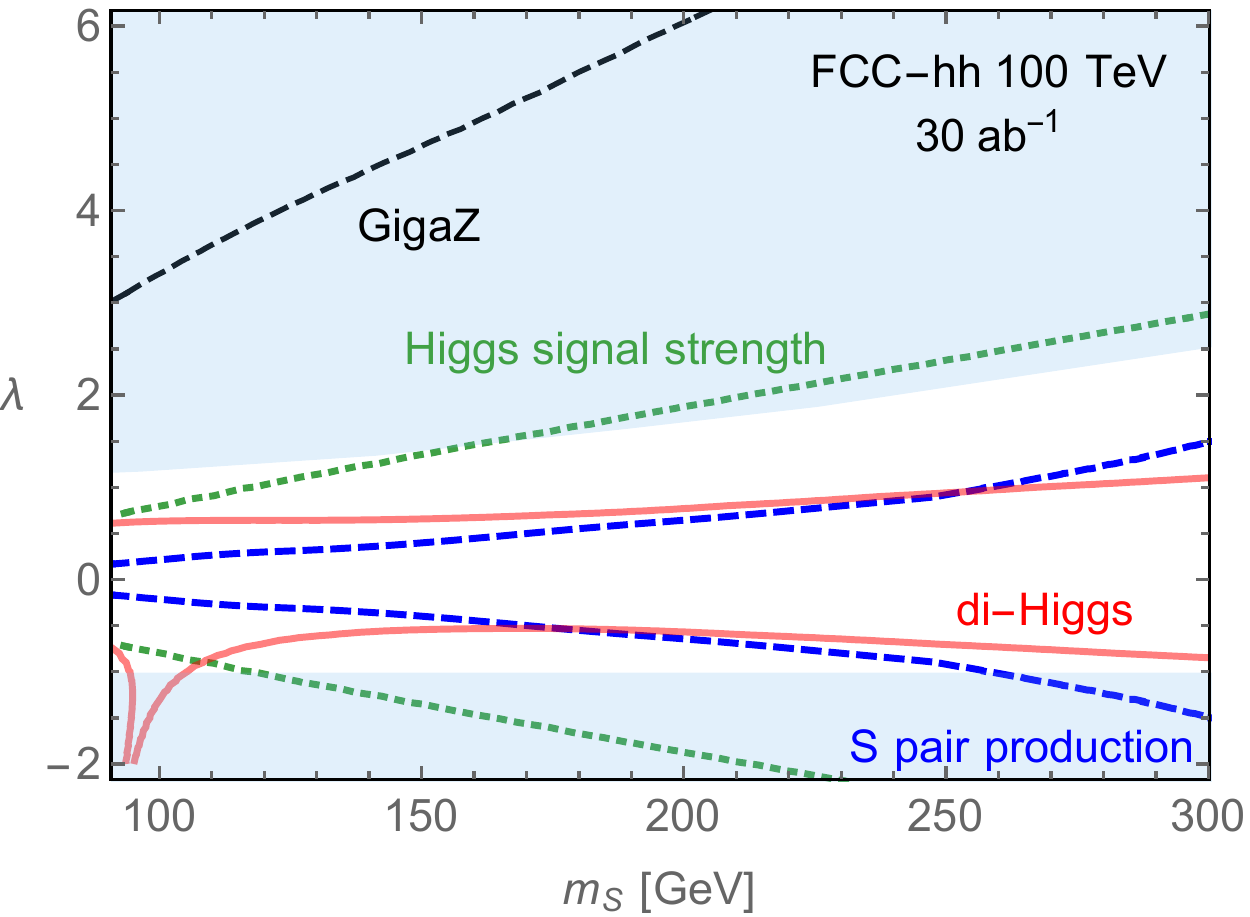}}
\hfill
\subfigure[\label{figlcbc}]{\includegraphics[width=8.3cm]{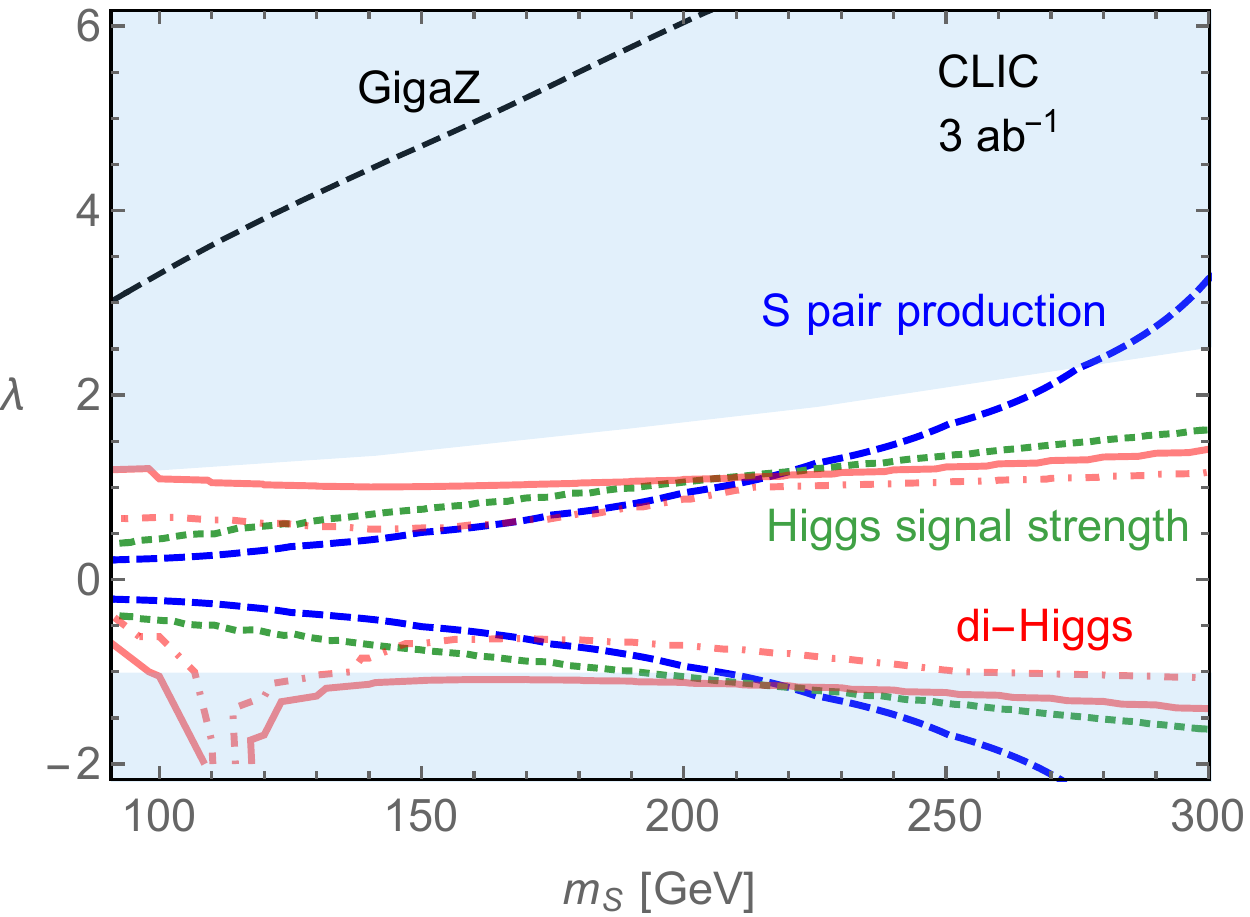}}
\caption{\label{fig:compariso} Comparison of the (a) FCC-hh and (b) CLIC at 3 TeV best case sensitivities. The dashed blue line refers to WBF to the sensitivity of $S$ pair production, while the dotted green line reflects the expected Higgs signal strength constraints.
For comparison we include the GigaZ extrapolation of the sensitivity via oblique corrections as a black dashed line. The projected self-coupling extraction are shown in red. For the 3 TeV CLIC the $5\%$- and $10\%$-level are included as dash-dotted and solid lines, respectively. For the Higgs signal strength here we use the 3~TeV value  $\delta\mu/\mu = 0.39\%$ according to Ref.~\cite{deBlas:2019rxi}. The shaded region is excluded from stability considerations~\cite{Curtin:2014jma} and assumes no additional BSM matter content beyond the singlet scalar. }
\end{figure*}

\section{Discussion and Conclusions}
\label{sec:conc}

We can now turn to a comparison of the expected sensitivity to this difficult to access new physics scenario. An overview for the examples of FCC-hh and CLIC is shown in Fig.~\ref{fig:compariso}. 

Amongst the indirect, loop-induced production processes di-Higgs production provides the best sensitivity, owing to the fact the $gg\to hh$ is largely driven by top-related threshold effects that are particularly sensitive to modifications of the trilinear Higgs coupling as predicted in the model at the expected self-coupling extraction precision. While for FCC-hh measurements of the Higgs signal strength cannot compete, the situation is different for lepton colliders such as CLIC, Fig.~\ref{fig:compariso}. For CLIC the Higgs self-coupling and the signal-strength measurements are comparable over a wide mass range. Off-shell $gg\to H \to ZZ$ production is essentially blind to this scenario due to gauge-related cancellations in the reasonable $\lambda$ range (cf. Fig.~\ref{fig:gg_ZZ}).

The electroweak precision constraint that we present here for the first time originate from the a priori most precise observables that can be obtained at the discussed colliders (see also the related Ref.~\cite{Chen:2020xzx} for recent theoretical developments). Unfortunately, due to the fact that the contributions from the new scalar only arise at two-loops, they will only start to become sensitive to $S$-induced modification if the expected sensitivity is improved by an order of magnitude .

At masses $\lesssim (200-250)\,{\rm GeV}$ the best sensitivity arises in $H\to SS$ off-shell missing energy searches. Together with Higgs-self-coupling measurements and, in the case of CLIC, the Higgs signal-strength measurements they are therefore the most promising avenues to discover or constrain the presence of weakly coupled scalars as expressed in Eq.~\eqref{eq:model}. If even more precise Higgs signal-strength measurements, as indicated by the FCC-ee projections, are achieved this could allow us to reach a similar sensitivity as the combined direct production and indirect di-Higgs probes at FCC-hh, as can be seen from Fig.~\ref{fig:higgssignal}. 

However, the importance of the different analyses (and the different collider concepts as a consequence), does crucially rest on the expected precision and control of the different final states measurements. For example, a 3\% accuracy of the Higgs self-coupling at a FCC-hh as detailed in Ref.~\cite{Contino:2016spe} provides favourable constraints at larger masses in the light of expected direct sensitivity. While this precision seems attainable from an experimental perspective (b-tagging, fake rates, etc.), it relies on an improvement of the theoretical uncertainty budget. Relaxing the self-coupling extraction to  $\sim 10\%$, direct off-shell $H\to SS$ limits start to dominate the sensitivity. As noted before, these do also suffer from small signal-over-background ratios and crucially depend on the understanding of the backgrounds. In both instances, data-driven techniques as considered in \cite{Aad:2016zqi,Aaboud:2018eqg,ATLAS:2020aub} could help to control uncertainties when perturbative improvements are out of reach. However, the combination of both channels capture the induced modifications over a wide range of masses.
At future lepton colliders, we find a similar picture.  The extrapolations of \cite{Charles:2018vfv,deBlas:2018mhx} suggest that the Higgs self-coupling can be determined at 3~TeV in the WBF at the 5-10\% level. Recast to the singlet scenario, we see that the di-Higgs production provides a slightly enhanced sensitivity at larger masses compared to direct and signal strength measurements. We note that this sensitivity crucially relies on the expected self-coupling precision. For instance, the slightly more conservative estimate of 22\% reported in Ref.~\cite{Abramowicz:2016zbo} is already too low to be competitive with the expected Higgs signal strength constraints.

While the scenario that we consider in this work is, by construction, difficult to observe, our results do suggest that the discovery potential of the FCC-hh concept can be similar or larger than that of lepton colliders in case systematics are under control. While this is not a surprise for heavy strongly-coupled physics such as SUSY, the combination of energy coverage and statistics, makes a naively sensitivity-limited hadron-hadron machine also an excellent tool to constrain weakly coupled electroweak extensions. In this sense, when power is applied in a controlled way to the symmetric Higgs portal, it will likely beat precision.

\noindent {\textbf{Acknowledgements}} --- 
We thank Federica Fabbri and Jay Howarth for helpful discussions.
C.E. is supported by the UK Science and Technology Facilities Council (STFC) under grant ST/P000746/1 and by the IPPP Associateship Scheme.
J.J. would like to thank the IPPP for hospitality and gratefully acknowledges support under the DIVA fellowship program of the IPPP. This work was completed while J.J. was at the Munich Institute for Astro- and Particle Physics (MIAPP) which is funded by the Deutsche Forschungsgemeinschaft (DFG, German Research Foundation) under Germany's Excellence Strategy – EXC-2094 – 390783311.
M.S. is supported by the STFC under grant ST/P001246/1. P.S. is supported by an STFC studentship under grant ST/T506102/1.

\bibliography{paper.bbl} 

\end{document}